\documentclass[fleqn,10pt]{wlscirep}

\usepackage{mathtools}
\usepackage{amssymb}
\usepackage{graphicx}
\usepackage{gensymb}
\usepackage{pdfpages}
\usepackage{enumitem}
\usepackage{pifont}
\usepackage{MnSymbol}
\usepackage{hyperref}
\usepackage{epsfig} 

\DeclarePairedDelimiter\bra{\langle}{\rvert}
\DeclarePairedDelimiter\ket{\lvert}{\rangle}

\DeclarePairedDelimiterX\braket[2]{\langle}{\rangle}{#1 \delimsize\vert #2}

\title{The phase-separation mechanism of a binary mixture in a ring trimer}

\author[1]{Vittorio Penna}
\author[1,*]{Andrea Richaud}

\affil[1]{Department of Applied Science and Technology and u.d.r. CNISM, Politecnico di Torino, I-10129 Torino, Italy}
\affil[*]{andrea.richaud@polito.it}

\begin{abstract}
We show that, depending on the ratio between the inter- and the intra-species interactions, a binary mixture trapped in a three-well potential with periodic boundary conditions exhibits three macroscopic ground-state configurations which differ in the degree of mixing. Accordingly, the corresponding quantum states feature either delocalization or a Schr\"odinger cat-like structure. The two-step phase separation occurring in the system, which is smoothed by the activation of tunnelling processes, is confirmed by the analysis of the energy spectrum that collapses and rearranges at the two critical points. In such points, we show that also Entanglement Entropy, a quantity borrowed from quantum-information theory, features singularities, thus demonstrating its ability to witness the double mixining-demixing phase transition. The developed analysis, which is of interest to both the experimental and theoretical communities, opens the door to the study of the demixing mechanism in complex lattice geometries.

\end{abstract}
\begin{document}

\flushbottom
\maketitle
\thispagestyle{empty}

\section*{Introduction}
Spatial phase separation of the atomic species forming a binary bosonic mixture
represents a rather intuitive phenomenon in which a sufficiently strong inter-species
repulsion is able to localize the two components in separated domains thus breaking the
spatial symmetry of the system.
This phase transition has attracted considerable attention \cite{cmixt1,cmixt2,cmixt3,cmixt4,cmixt5} 
within the physics of binary condensates and the mechanism of phase separation,
investigated in the condensate mean-field picture, has revealed a nontrivial
dependence from trapping potentials, the boson number of each species and other significant
parameters of the system. In parallel, dynamical stability and the emergence of excited states of binary 
mixtures have been thoroughly studied in the transition regime to
the immiscible phase \cite{kasa,Gallemi_1,tic,lee}. 

The realization of optical lattices \cite{jz,bdz,yuk} and the trapping of binary mixtures therein\cite{Inguscio,Gadway,Soltan} have further stimulated the theoretical
study of mixtures with the aim to understand their properties when the inter-species coupling combines with 
the effect of spatial fragmentation of the two species in the potential wells forming
the lattice.
An extraordinarily rich scenario has resulted 
from this work which, in addition to phase separation \cite{sep1,sep2,sep3, Angom}, has explored 
the emergence of exotic magnetic-like phases \cite{ks,ddl} and quantum emulsions 
\cite{qe1}, 
the deformation of Mott domains due to the presence of a second species \cite{mott}, 
the formation of polaron excitations \cite{pol}, the inter-species entanglement \cite{ent},
the modulation instability in the separation regime \cite{modul} and
the excitation of persistent currents in a ring lattice \cite{NoiPRA2}.

Recently, the spatial phase separation of a binary quantum fluid has been investigated
by considering two bosonic species trapped in the simplest two-well potential. The resulting model, 
a two-species Bose-Hubbard Hamiltonian, allows one to perform a fully analytic derivation 
of the critical condition characterizing the space-localization effects \cite{PennaLinguaJPB,PennaLinguaPRE}. 
In addition to highlight how phase separation is marked by the spectral collapse of energy levels (an effect already observed for spin chains and dipolar bosons in a trimer \cite{Gallemi_3,Gallemi_4}), this 
study reproduces, in the presence of large boson numbers, the well known critical value $W/U =1$ 
(which links the inter-species interaction $W$ and the intra-species interaction of twin species 
with $U=U_a=U_b$) characterizing the spatial separation in large-size lattices at zero 
temperature \cite{sep3}. 
Apparently,   
neither the space fragmentation of the two species among the potential wells representing the
lattice sites nor the size of the lattice significantly affect the spatial phase separation
expected for $W/U>1$.

Unexpected features are found to characterize the species transition to the demixed state when considering 
a binary fluid at zero temperature whose two components are distributed in a three-well potential. This becomes evident when the interspecies interaction, the control parameter driving demixing, 
is varied. The three-well system reveals a separation mechanism totally different from that
observed in a binary mixture distributed in a two-well potential.
The nontrivial character of the transition from the regime with 
totally-mixed components to the regime with fully-separated components is that such regimes 
are intercalated by an intermediate phase where a partial phase separation takes place 
in two of the three wells while in the remaining one the two components are still completely mixed. 
Only a further increase of the interaction ratio $W/U$ is able to remove the residual mixing
leading to a complete space separation. Space fragmentation of the two components in 
three wells thus have a considerable effect on their separability properties. 

The triple-well system in a ring geometry, essentially consisting in a three-site Bose-Hubbard model, has attracted large attention due to the rich scenario of dynamical behaviours including instabilities\cite{PBF}, vortex states\cite{Jason}, coherent\cite{Bradly} and $\pi-$phase \cite{Gallemi_2} tunnelling. 
The model describing a binary mixture in a triple well is effectively represented by two Bose-Hubbard Hamiltonians, each one associated to a single component and depending on three spatial boson modes, and by the density-density interspecies coupling of the two components.
This has the form
\begin{equation}
\label{eq:Hamiltoniana}
  \hat{H}= - T_a \sum_{j=1}^{3} \left(A_{j+1}^\dagger A_j +A_j^\dagger A_{j+1} \right) + \frac{U_a}{2} \sum_{j=1}^{3} N_j(N_j-1) - T_b \sum_{j=1}^{3} \left(B_{j+1}^\dagger B_j +B_j^\dagger B_{j+1} \right)
  +\frac{U_b}{2} \sum_{j=1}^{3} M_j(M_j-1)+W \sum_{j=1}^{3} N_j\, M_j 
\end{equation}
where $j=4 \equiv 1$ due to the periodic boundary conditions, $T_a$ and $T_b$ are the hopping amplitudes, $U_a$ and $U_b$ the \textit{intra}-species repulsive interactions, and $W$ represents the \textit{inter}-species repulsion. $A_j$ and $B_j$ are standard bosonic operators, having commutators $[A_j,A_k^\dagger]=\delta_{j,k}=[B_j, B_k^\dagger]$, and $[A_j,B_k^\dagger]=0$. After defining number operators $N_j=A_j^\dagger A_j$ and $M_j=B_j^\dagger B_j$, we recall that total boson numbers $N = \sum_j N_j$ and $M = \sum_j M_j$ of both species represent conserved quantities, namely, $[\hat{H}, N]= [\hat{H},M]=0$. We shall consider a mixture with twin condensates where $U_a=U_b=U$ and $T_a=T_b= T$. Slight deviations $U_a\ne U_b$ and $T_a\ne T_b$ from this ideal case, as one expects to observe in real systems, can be shown to not alter in a significant way our results.    

\section*{Results}
The study of the equilibrium configurations of a bosonic binary mixture in a three-well potential (trimer), reveals how, according to the ratio between the interspecies and the intraspecies repulsion, $W/U$, three different families of qualitatively different ground states emerge in the zero-temperature regime. The main effect analysed in this paper, the two-step process where different mixing-demixing phase transitions are triggered by a change in $W/U$, is a) investigated by means of a semiclassical approach to determine the ground-state configurations, b) shown to be related to the collapse and rearrangement of the energy spectrum, c) characterized in terms of quantum-correlation properties (entanglement entropy) between two parts of the systems.

\paragraph*{Demixing transitions.}
It is advantageous to investigate spatial phase separation by reformulating Hamiltonian (\ref{eq:Hamiltoniana})
within the continuous variable picture (CVP) described in the section Methods. This allows one to replace $\hat H$
with an effective Hamiltonian operator typically providing a satisfactory representation of low energy states.     
In particular, the CVP approach allows one to reduce the search for the ground state of Hamiltonian (\ref{eq:Hamiltoniana}) to the one for the minimum of the following effective potential:
$$
V_*=  -2NT\left(\sqrt{x_1x_2}+\sqrt{x_2x_3}+\sqrt{x_3x_1}+\sqrt{y_1y_2}+\sqrt{y_2y_3}+\sqrt{y_3y_1}\right)
  + 
$$
\begin{equation}
\label{eq:V_*}
  \frac{UN^2}{2}\left(x_1^2+x_2^2+x_3^2+ y_1^2+y_2^2+y_3^2 \right)+ WN^2 \left(x_1y_1+x_2y_2+x_3 y_3 \right)
\end{equation}
where variables $x_i$ and $y_i$ ($i=1,2,3$) are normalized boson numbers, entailing that $x_i,\,y_i \in[0,1]$ and that $z_1+z_2+z_3=1$, ($z=x,\,y$) due to particle number conservation and $T_a=T_b=:T$, $U_a=U_b=:U$. Notice that only many-body states with the same fixed number $N=M$ of bosons for the two species are considered. The configuration $(\vec{x},\vec{y})$ which minimizes the effective potential $V_*$ is determined, at first, when the tunnelling $T$ is suppressed. In this situation, in fact, it is possible to carry out a fully analytic study (see Methods section), capable of illustrating the physics of the problem in a particularly simple and effective way. The results, sketched in Fig. \ref{fig:Popolazioni_T_nullo}, can be summarized as follows:

\begin{figure}[ht]
\centering
\includegraphics[width=\linewidth]{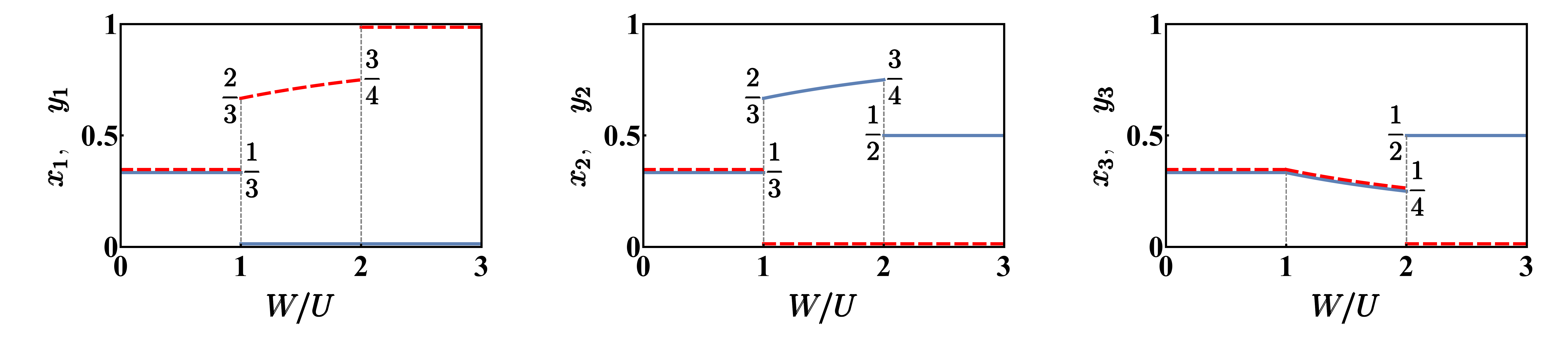}
\caption{Normalized semiclassical bosonic populations in each of the three wells for varying $W/U$ and for $T=0$. Each panel corresponds to a site of the trimer, while blue solid (red dashed) lines are associated to bosonic species a (b). (Notice that, wherever two lines overlapped, they have been slightly displaced for purely graphical reasons).  One can recognize the presence of three different quantum phases: i) for $W/U\in[0,1)$ the two condensates are fully mixed and completely delocalized; ii) for $W/U\in(1,2)$ they are separated in two out of the three wells while they coexist in the third one; iii) for $W/U>2$ the demixing is complete, meaning that one species conglomerates in a well, while the other species distributes in the remaining two wells. It is clear that $W/U=1,2$ represent critical values where two different mixing-demixing phase transitions occur. Data have been obtained by means of a fully analytic study of the global minimum of effective potential (\ref{eq:V_T_nullo}) (see Methods section).}
\label{fig:Popolazioni_T_nullo}
\end{figure}

\begin{enumerate}[label=(\roman*)]
    \item  For small $W/U$ values (more specifically, if $W/U \in [0,1)$), the two species are delocalized and uniformly distributed in the trimer, thus entailing a perfect mixing. 
    \item For moderate $W/U$ values (namely, if $W/U \in (1,2)$), the demixing occurs in two out of the three wells, while, in the third well, the species are still mixed. Of course, due to the symmetry of the trimer system, no well is favoured compared to the others, and, as a consequence, there are three configurations $(\vec{x},\vec{y})$ which minimize the effective potential $V_*$. Among such configurations, which are equal up to a cyclic permutation of the site labels, the one that we have illustrated in Fig. \ref{fig:Popolazioni_T_nullo} is
    \begin{table}[ht]
    \centering
    \begin{tabular}{ l l l| l l l| l l l}
        $x_1=0$ & & & & $x_2=\frac{1+\frac{W}{U}}{2+\frac{W}{U}}$ & & & & $x_3=\frac{1}{2+\frac{W}{U}}$ \\
        $y_1=\frac{1+\frac{W}{U}}{2+\frac{W}{U}}$ & & & & $y_2=0$ & & & & $y_3=\frac{1}{2+\frac{W}{U}}$ \\
    \end{tabular}
    \end{table}

   Quantum-mechanically, the degeneracy associated to the three minimum-energy configurations leads to the formation of a non-degenerate ground state $\ket{\Psi_0}$ which is a three-sided Schr\"odinger cat, i.e. a state of the type 
   \begin{equation}
       \label{eq:Cat_3}
       \ket{\Psi_0} = \frac{1}{\sqrt{3}}\sum_{j=1}^3 \ket{\psi_j}
   \end{equation}
   where each $\ket{\psi_j}$ corresponds to a different macroscopic configuration.
   Our analysis, which is semiclassical, breaks the symmetry of the problem and considers just one side of the cat, the one defined above and sketched in Fig. \ref{fig:Popolazioni_T_nullo}.    
    It is interesting to observe that, crossing the critical point $W/U=1$, the populations of site 1 and site 2 exhibit discontinuities, i.e. jumps from the uniform configuration $x_1=y_1=x_2=y_2=1/3$ to the characteristic values $2/3$ and $0$. Conversely, functions $x_3(W/U)$ and $y_3(W/U)$ are continuous in $W/U=1$. Increasing the ratio $W/U$ from $1$ to $2$, the third site (where the two bosonic species coexist) gradually looses bosons, while the first and the second site (each one hosting just one species) become more and more populated. For $W/U\to 2^-$, the first well hosts $3/4$ of species-b bosons, while the second well hosts $3/4$ of species-a bosons. The remaining parts coexist in the third well in equal measure.   
    
    \item For big $W/U$ values (i.e., for $W/U>2$), the two species completely demix. One of them conglomerates in one site, while the other one spreads on the remaining two sites. This situation corresponds to $6$ possible scenarios, i.e. to 6 configurations $(\vec{x},\vec{y})$ minimizing the effective potential $V_*$. In fact, if configuration $(\vec{x}^*,\vec{y}^*)$ is a minimum for $V_*$, also the two configurations obtained by cyclic permutations of the site labels are minima for $V_*$. Moreover, as the species we are considering feature the same hopping amplitude $T$ and the same on-site repulsion $U$, the configurations where species-a conglomerates must not be favoured to the configurations where it is species-b the one that conglomerates. In other words, the energy of the system remains equal after the exchanges $x_i \leftrightarrow y_i$ $(i=1,2,3)$. The presence of $6$ minimum-energy configurations in the classical analysis corresponds to a quantum ground state $\ket{\Psi_0}$ which is a six-faced Schr\"odinger cat, i.e. a state of the type
    \begin{equation}
        \label{eq:6-cat}
   \ket{\Psi_0}=\frac{1}{\sqrt{6}}\sum_{j=1}^6 \ket{\psi_j}
    \end{equation}
    where each $\ket{\psi_j}$ is a state corresponding to a certain macroscopic configuration.
    In Fig. \ref{fig:Popolazioni_T_nullo} we have broken such symmetry and plotted just one of the $6$ possible configurations.     
\end{enumerate}

\begin{figure}[ht]
\centering
\includegraphics[width=\linewidth]{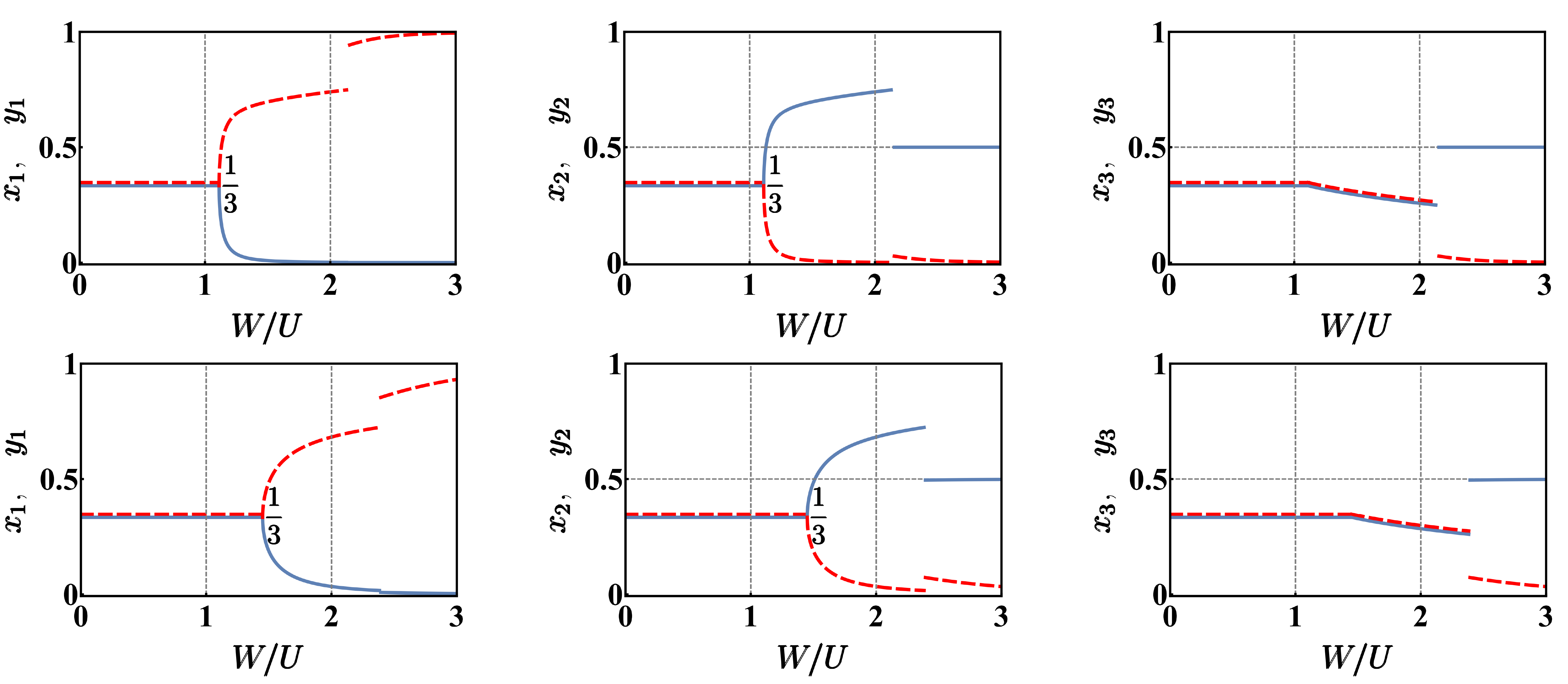}
\caption{Normalized semiclassical bosonic populations in each of the three wells for varying $W/U$ and for $N=50$. First row corresponds to $T/U=1.25$, second row to $T/U=5$; Each panel corresponds to a site of the trimer, while blue solid (red dashed) lines are associated to bosonic species a (b). The presence of a non-zero $T$ is responsible for the rightward translation of the two critical points. Data correspond to a particular branch of solutions (determined numerically) of system (\ref{eq:Sistema_4}).}
\label{fig:T_non_nulli_4}
\end{figure}

\bigskip
\noindent
As illustrated in Fig. \ref{fig:T_non_nulli_4}, for non-zero hopping amplitudes $T$, the critical points move rightward (i.e., the transitions occur at bigger values of $W/U$). 
It is possible to show with a fully analytic computation based on the request that Hessian matrix (\ref{eq:Hess}) must be positive definite (see Methods section), that the first mixing-demixing phase transition onsets at
\begin{equation}
\label{eq:critical_ration}
    \frac{W}{U}  = 1+ \frac{9}{2}\frac{T}{UN},
\end{equation}
a value which is greater than $1$, mirroring the fact that tunnelling $T$ favors the mixed phase and delays the occurrence of the first demixing. On the other hand, one can notice that the effect of $T$ is smaller and smaller as the number of bosons, $N$, increases and, in the thermodynamic limit, it becomes negligible. In such limit, the scenario is again the one depicted in Fig. \ref{fig:Popolazioni_T_nullo}.

\begin{figure}[ht]
\centering
\includegraphics[width=1\linewidth]{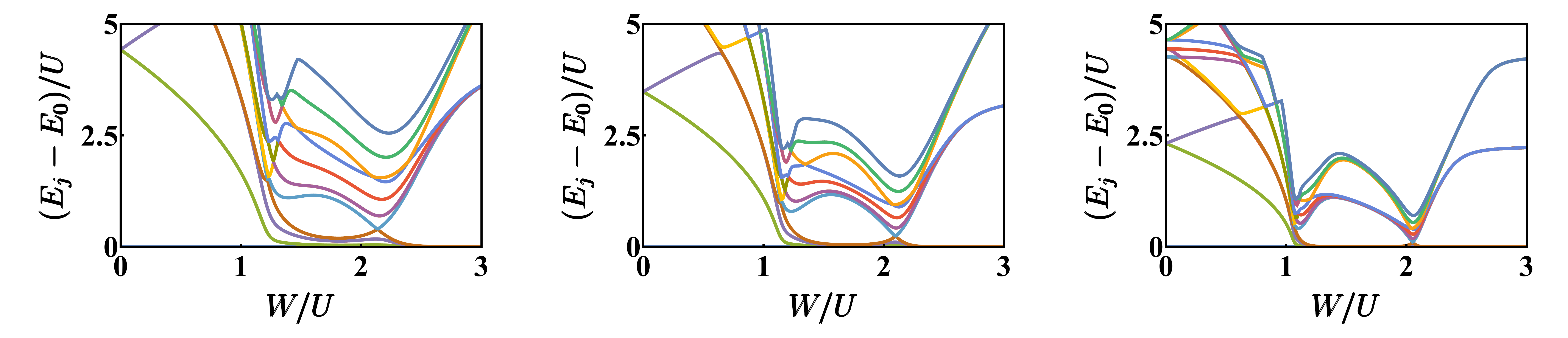}
\caption{First $16$ energy levels obtained for $N=15$, $T/U=0.6,\,0.4,$ and $0.2$ respectively. The smaller $T/U$, the more evident the rearrangement and the collapse of the spectrum at the critical points. Data have been obtained by means of an exact diagonalization of Hamiltonian (\ref{eq:Hamiltoniana}). Notice that the first panel (the one obtained for $T=0.6$) can be compared with boson populations depicted in Fig. \ref{fig:Most_probable}.}
\label{fig:Trimer_spectral_collapse}
\end{figure}

\paragraph{Collapse and rearrangement of the energy spectrum.} 
The dramatic change in the ground state qualitative structure occurring at the two critical values of $W/U$ is mirrored by a deep rearrangement of the energy spectrum across these two transition points. As shown in Fig. \ref{fig:Trimer_spectral_collapse}, the energy levels (obtained by means of an exact diagonalization of Hamiltonian (\ref{eq:Hamiltoniana})) tend to collapse in the neighborhood of the critical values, meaning that the energy cost to create excitations (or, more generally, the interlevel distance) tends to vanish. This effect is more evident increasing the number of bosons $N$ and/or decreasing the hopping amplitude $T$.        

\begin{figure}[ht]
\centering
\includegraphics[width=0.5\linewidth]{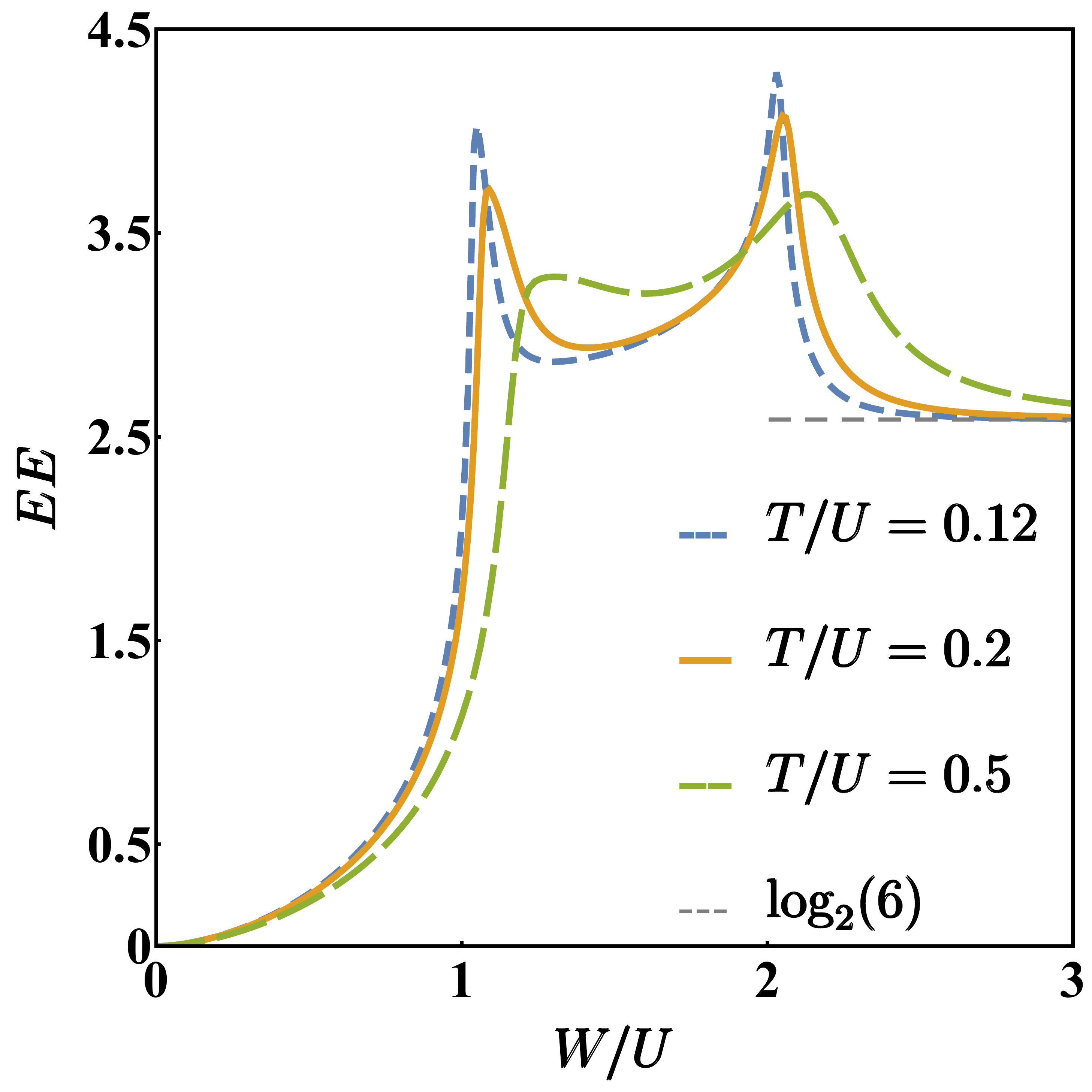}
\caption{Entanglement entropy between the two atomic species for different values of $T/U$ and for $N=15$. Notice that, the smaller the ratio $T/U$, the sharper the peaks. Data have been obtained by means of an exact diagonalization of Hamiltonian (\ref{eq:Hamiltoniana}).}
\label{fig:Trimer_entanglement_A_B}
\end{figure}

\paragraph{Entanglement entropy as a critical indicator.} As is well known, the entanglement entropy (EE) quantifies the quantum correlation between two parts of a physical system through the Von Neumann entropy of a suitably defined sub-system\cite{NoiEntropy,RevModPhysAmico}. In this work, the physical system is of course the binary mixture, the two parts correspond to the two different atomic species and, therefore, the EE describes the quantum correlation between them (see Methods section for a detailed mathematical definition). As shown in Fig. \ref{fig:Trimer_entanglement_A_B}, this indicator strongly depends on the specific phase of the system which, in turn, is determined by the ratio $W/U$. 
In the completely mixed phase, i.e. for $W/U$ smaller than the critical value ($\ref{eq:critical_ration}$), the EE steadily increases as $W/U$ increases, despite the fact that semiclassical boson populations remain constant (uniformly distributed on the three wells). Notice that EE $\to 0$ for $W/U\to0$ because, in this limit, the species are decoupled. At the two critical values of $W/U$, the EE features peaks, which not only indicate that the two mixtures are strongly entangled, but also allow to clearly distinguish the three different phases. Such peaks are sharper if $T/U$ is smaller, that means that the effect of the hopping amplitude is to smooth the mixing-demixing transition. For $W/U$ greater than the second critical value, the demixing gets more and more complete and the EE approaches the limiting value $\log_2(6)$. According to standard notions of Quantum Information Theory and to applications thereof in the field of ultracold bosons \cite{Gallemi_2,NoiEntropy}, the emergence of this value can be appreciated by recalling that there are $6$ possible ways of realizing a completely demixed semiclassical configuration which, in turn, equally contribute to the formation of a unique quantum ground state of the type (\ref{eq:6-cat}).


\section*{Discussion}
We have investigated the two-step phase separation of a binary mixture in a trimer, highlighting the presence of an intermediate phase, which is neither completely mixed nor completely demixed. This element originates from the fact that the number of sites is odd and it is absent in the two-well system \cite{PennaLinguaPRE,PennaLinguaJPB}, where there is a one-step transition from a fully mixed to a fully demixed configuration. 

The problem of determining the ground state has been reduced, by means of the CVP approach (see Methods section), to the search for the minimum-energy configuration of effective potential (\ref{eq:V_*}). The latter features an effective energy landscape whose minimum points are formed or disappear depending on the ratio $W/U$. We start analyzing the situation where the tunnelling $T$ is suppressed. This case is of particular importance not only because it is associated with fully analytic results, but also because it captures the system behavior in the large $N$ limit. As explained in the Methods section, the search for the global minimum of function (\ref{eq:V_*}) is not trivial, as one has to take into account that the \textit{global} minimum is not necessarily a \textit{stationary point}. A careful analysis, based on a systematic exploration of the domain boundary leads to the identification of three different phases, which qualitatively differ in the degree of mixing and spatial localization (see Fig. \ref{fig:Popolazioni_T_nullo}). 

\begin{figure}[ht]
\centering
\includegraphics[width=\linewidth]{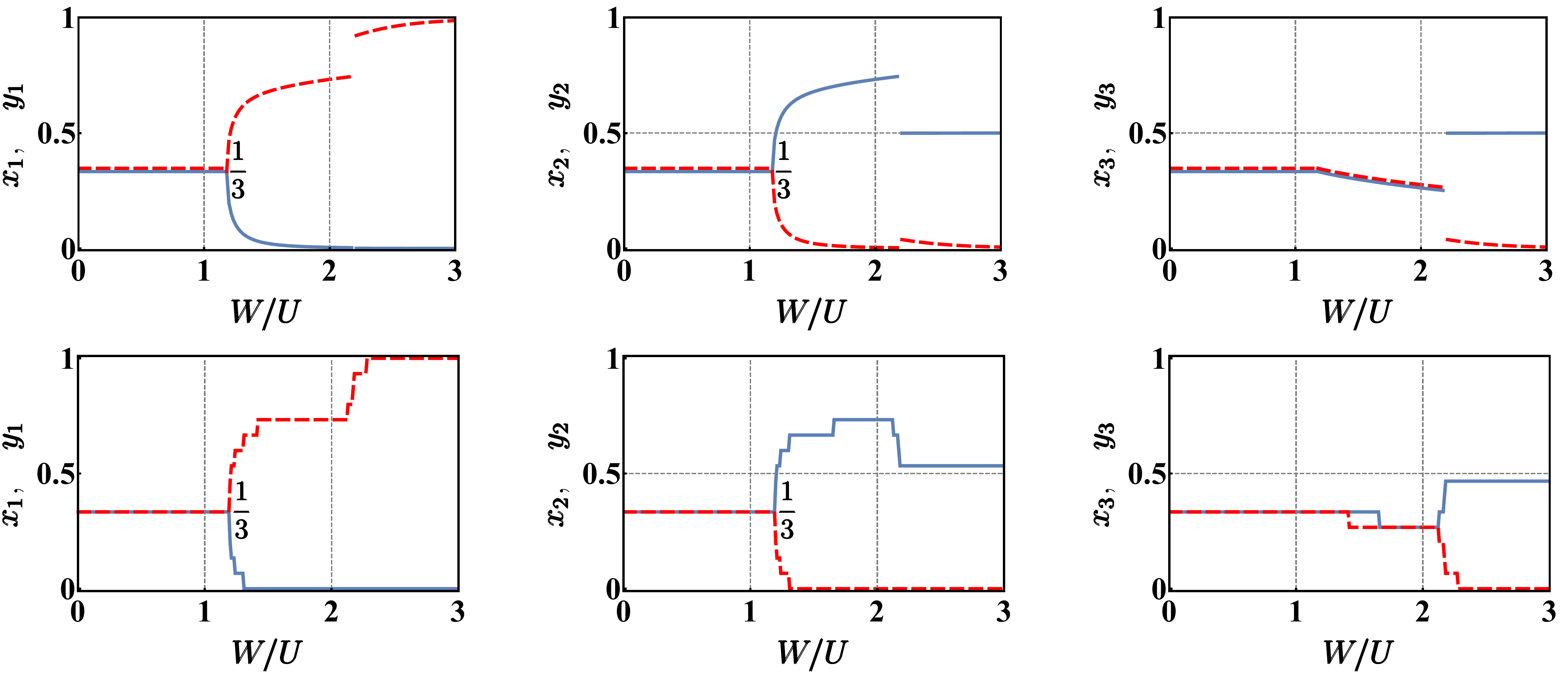}
\caption{Upper row: a branch of solutions predicted by the CVP approach. Lower row: the corresponding branch of most probable Fock states present in the ground states $\ket{\Psi_0}$. The latter have been determined by means of an exact numerical diagonalization. In both cases parameters $N=15$, $U=1$, $T=0.6$ have been chosen. Each panel corresponds to a site of the trimer, while blue solid (red dashed) lines are associated to bosonic species a (b).}
\label{fig:Most_probable}
\end{figure}

The activation of the hopping amplitude $T$ makes the model more realistic but does not distorts the scenario, which is still marked by two critical points. From a technical point of view, the search for the minimum energy configuration is simpler because $T$ constitutes a regularizing term which prevents the points of minimum from falling on the boundary. The exploration of the domain boundary is therefore no longer necessary and it's enough to look just for \textit{local} points of minimum in the interior of the domain. This conceptual simplification comes with the impossibility of analytically solving the resulting system of equations (\ref{eq:Sistema_4}).  With reference to Fig. \ref{fig:T_non_nulli_4}, where numerical results are plotted, one can observe that bigger $W/U$ values are needed to trigger the demixing transitions which, in turn, are less abrupt. To test their reliability, we have compared CVP's predictions to the results obtained by means of exact numerical diagonalization.  Fig. \ref{fig:Most_probable} shows that the two mixing-demixing phase transitions occur at the same critical values of $W/U$ and, more generally, that the semiclassical approach and the purely quantum scenario are in good agreement. Notice that we chose not to plot the expectation values of boson populations in the ground state (e.g. $\bra{\Psi_0}N_j\ket{\Psi_0}$) because one would obtain the \textit{average} over the different macroscopic configurations constituting the cat-states and, therefore, the constant value $N/3$, irrespective of the ratio $W/U$. Instead, we have determined, sweeping the parameter $W/U$, the \textit{most probable configuration} i.e. the Fock state associated to the biggest (square modulus of the) coefficient in the decomposition of the ground state.
It is interesting to notice that function (\ref{eq:V_*}) features a unique minimum in the uniform and fully mixed phase, three isoenergetic points of minimum in the intermediate phase and six isoenergetic points of minimum in the fully demixed phase. These degeneracies in the semi-classical energy landscape are resolved, quantum mechanically, with the formation of cat-states of the type (\ref{eq:Cat_3}) and (\ref{eq:6-cat}) respectively.

\begin{figure}[ht]
\centering
\includegraphics[width=\linewidth]{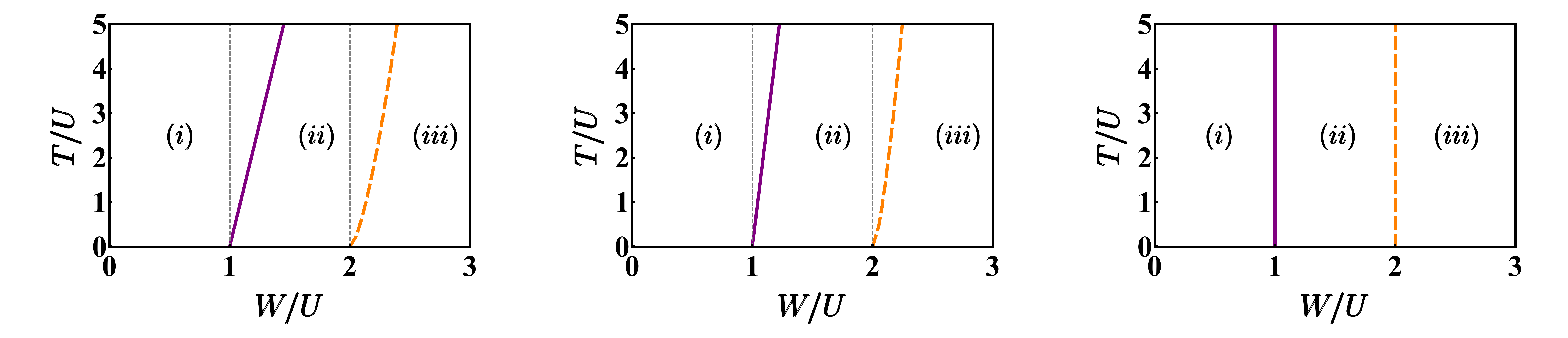}
\caption{Phase diagrams for $N=50$, $N=100$, $N\to +\infty$. Purple solid lines correspond to critical condition (\ref{eq:critical_ration}); orange dashed lines corresponds to those solutions of system (\ref{eq:System_2_equations}) which are no longer minimum points for (\ref{eq:V_*}). Label (i) indicates the uniform and fully mixed phase, region (ii) is the intermediate phase and, eventually, label (iii) denotes the fully demixed phase. Notice that, as $N$ increases, the role of $T$ becomes more and more negligible. }
\label{fig:Diagramma_di_Fase}
\end{figure}

The phase diagram of the binary mixture on a three-well potential, sketched in Fig. \ref{fig:Diagramma_di_Fase}, has been determined numerically solving systems (\ref{eq:Sistema_4}) and (\ref{eq:System_2_equations}), and checking that the obtained stationary point actually is a minimum point for effective potential (\ref{eq:V_*}). Purple solid lines represent critical condition (\ref{eq:critical_ration}) derived analytically while no analytic expression can be found to define the orange dashed lines due to the complex nonlinear character of equations (\ref{eq:Sistema_4}). These lines have been plotted on the basis of numerical data (see Methods section for details). 

\begin{figure}[ht]
\centering
\includegraphics[width=\linewidth]{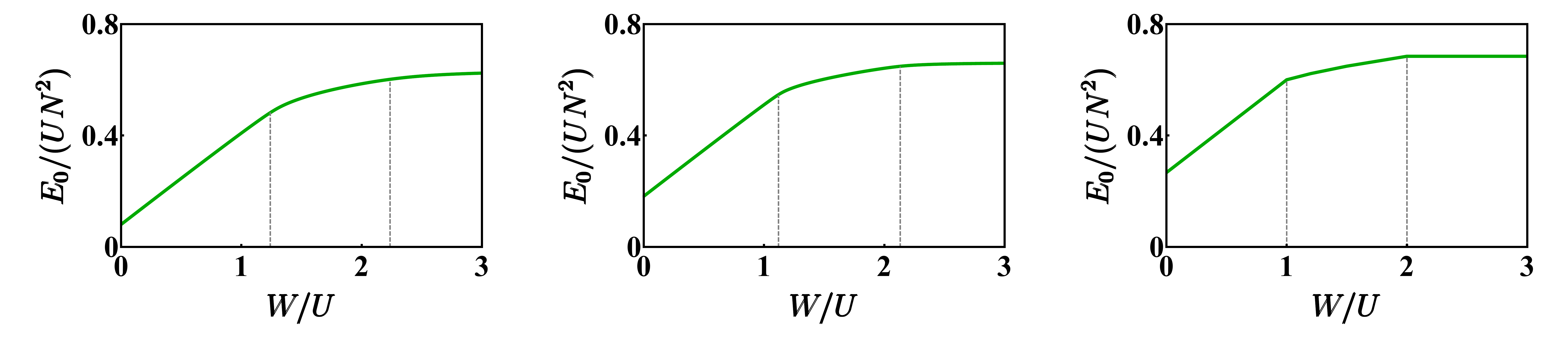}
\caption{Normalized ground state energy vs $W/U$ for $N=15$ bosons. In each panel, the ratio $T/U$ has been set to $0.8$, $0.4$ and $0$ respectively. One can observe that, for non-zero $T$ (first and second panel), the function is differentiable everywhere, while, for $T/U \to 0$ (third panel), the function is not differentiable at the two critical points. Gray dashed lines have been drawn where the CVP approach predicts the occurrence of mixing-demixing phase transitions (notice that they move leftward for $T/U\to 0$). Data have been obtained by means of an exact diagonalization of Hamiltonian (\ref{eq:Hamiltoniana}).}
\label{fig:Andamento_E}
\end{figure}

The double critical behaviour exhibited by the system upon a variation of $W/U$ can be evidenced also from the energy-spectrum standpoint \cite{PennaLinguaPRE}. As shown in Fig. \ref{fig:Trimer_spectral_collapse}, whose data have been obtained by means of an exact (numerical) diagonalization of Hamiltonian (\ref{eq:Hamiltoniana}), the energy levels tend to collapse in the neighborhood of the two critical points and have different arrangements in the three phases. One can recognize, for example, families of energy levels which are clearly distinct in the first two phases while tend to overlap in the third one. Approaching the limiting case $T\to 0$ (note that this can be shown to be equivalent to taking $N\to +\infty$), the collapses are more and more evident and the points where they take place tend to $W/U=1$ and $W/U=2$. The low-energy excitations spectrum of a binary mixture in the uniform and completely mixed phase was determined for a generic $L-$site ring by means of a group-theoretic approach\cite{NoiPRA2} combined with Bogoliubov scheme. Choosing $L=3$ sites, the condition corresponding to the collapse of a Bogoliubov frequency exactly reproduces equation (\ref{eq:critical_ration}), which gives the onset of the first mixing-demixing phase transition. One can thus appreciate the fact that the range of validity of the quasi-particles Bogoliubov spectrum exactly corresponds to the region of parameters space where the system is in the uniform and fully mixed phase. 
This analysis is confirmed by the behaviour of the ground state energy $E_0=\bra{\psi_0}\Hat{H}\ket{\psi_0}$ of Hamiltonian (\ref{eq:Hamiltoniana}). $E_0$ has been numerically evaluated as a function of parameter $W/U$, for different values of $T/U$. The results, illustrated in Fig. \ref{fig:Andamento_E}, show, for $T/U\to 0$ (or, equivalently, for $N\to+\infty$), the emergence of two critical points at $W/U=1$ and $W/U=2$. In such points, the ground state energy features discontinuities in the first derivative, a circumstance which suggests the onset of a quantum phase transition. 
The systematic study of this double phase transition will be developed elsewhere in the same spirit of the localization-delocalization transition in the Bose-Hubbard trimer\cite{BuonsantePennaVezzani}. 

Quantities traditionally belonging to quantum information theory have been used to highlight the occurrence of quantum phase transitions\cite{Mazz, DellAnnaMazzarella, Hines, FuLiu, JuliaDiaz, Viscondi}. Among the others, Entanglement Entropy has proved to be particularly effective in detecting the quantum criticality of the localization transition in a bosonic Josephson junction\cite{BuonsanteBurioni}. This indicator, which measures the degree of entanglement between two parts of a given system through the Von Neumann entropy of a reduced subsystem, has been shown to be sensible also to the mixing-demixing phase transition occurring in a two-species bosonic mixture loaded in a dimer\cite{PennaLinguaJPB} and to be robust with respect to the change of partition scheme \cite{NoiEntropy}. As explained in the Methods section, one starts from the density matrix associated to the ground state, i.e. from $\hat{\rho}_0=\ket{\psi_0}\bra{\psi_0}$. Then, a partition scheme aimed at identifying two sub-systems is chosen. In our case one sub-system corresponds to species-a bosons while the other one corresponds to species-b bosons. The density matrix of a reduced subsystem, $\hat{\rho}_{0,a}=\mathrm{Tr}_b\left( \hat{\rho}_0\right)$, is obtained by tracing out the degrees of freedom of the other one (see formula (\ref{eq:Reduced_density_matrix})). One thus remains with the reduced density matrix of a mixed state and the Von Neumann entropy thereof (\ref{eq:Von_Neumann_Entropy}) corresponds to the entanglement between the two sub-systems, namely between the two bosonic species. 

By means of the exact (numerical) diagonalization of Hamiltonian (\ref{eq:Hamiltoniana}), we have computed the EE as a function of $W/U$ for different values of $T/U$. The results, illustrated in Fig. \ref{fig:Trimer_entanglement_A_B} and discussed in the previous section, not only demonstrate the ability of EE to capture the double critical behaviour of the system upon a variation of $W/U$, but also witness the formation of 6-faced cat-like states of the type (\ref{eq:6-cat}) for large $W/U$ values. Notice, in fact, that, for $W/U\to +\infty$, the EE tends to the limiting value $\log_2(6)$, where the number $6$ is the number of different semiclassical configurations $(\vec{x},\,\vec{y})$ which minimize effective potential (\ref{eq:V_*}) when $W/U\gg 3$. 

\begin{figure}[ht]
\centering
\includegraphics[width=0.6\linewidth]{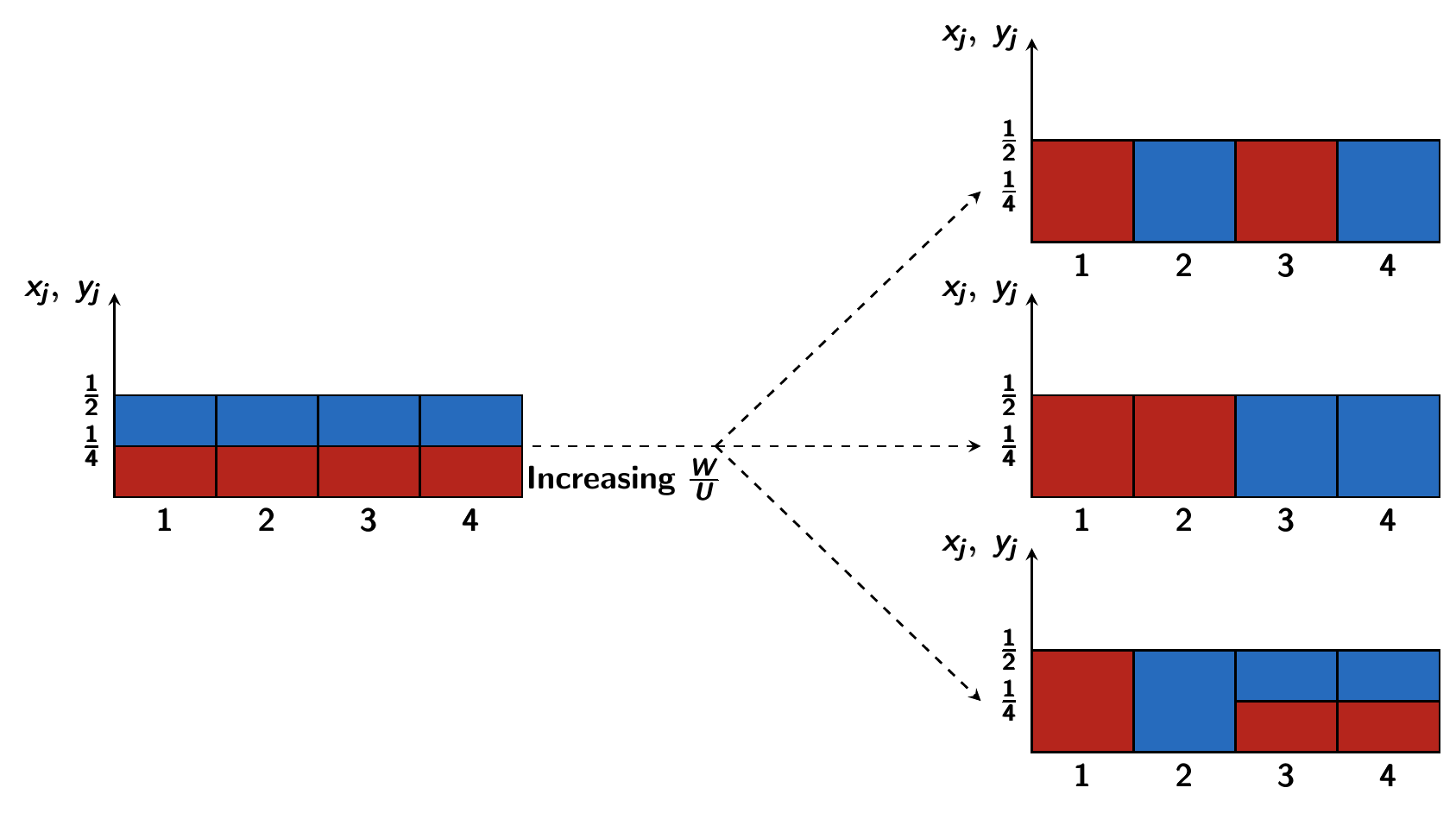}
\caption{Possible phase-separation mechanisms for a binary mixture in a $4$-well potential with periodic boundary conditions. Blue (red) color corresponds to species a (b). On the left, the fully mixed configurations is depicted. On the right, some examples of demixed configurations (from top to bottom: complete demixing with emulsion-like structure, complete demixing with well separated structure, and partial demixing). Left panel corresponds to $W/U<1$, right panels to $W/U=1$. $T/U=0$ has been chosen for all plots.}
\label{fig:Ipotesi_4_buche}
\end{figure}

In this work, we have evidenced a phase separation mechanism considerably more complex than the one occurring in a two-well system \cite{PennaLinguaPRE,PennaLinguaJPB,NoiEntropy}. The presence of an intermediate phase, which stands between the uniformly mixed configuration and the completely demixed phase, opens the door to the study of the phase separation mechanism in rings having an even or an odd number of sites and, more generally, in more complex lattice topologies. An illustrative example is shown in Fig. \ref{fig:Ipotesi_4_buche}, where we consider a $4$-well ring lattice. The CVP approach leads to a generalization of effective potential  (\ref{eq:V_*}) which, in the case of vanishing tunnelling and at the transition point $W/U=1$, can be shown to be minimized by any macroscopic configuration of the type $y_j=1/2-x_j$, with $j=1,2,3,4$ (four of them are shown in Fig. \ref{fig:Ipotesi_4_buche}). As soon as $W/U>1$, this degeneracy is broken and one macroscopic configuration gets energetically more favorable. The determination a) of the ground state configuration when $W/U>1$ and for a generic number of sites $L>3$, b) of possibly manifold intermediate phases and c) the exploration of the potential-$V_*$ domain represented by high-dimensional polytopes (see the section Methods) indeed constitute complex problems which require a separate extended analysis. 
Their solution will be addressed in a future work.          

\section*{Methods}
\paragraph{The continuous variable picture.} 
An effective study of the ground state configuration of multimode BH Hamiltonians can be carried out by reformulating single-site populations $N_j$ and $M_j$ in terms of continuous variables $x_j=N_j/N$ and $y_j=M_j/M$ \cite{BuonsantePennaVezzani,PennaLinguaPRE,PennaFranzosi2001,Spekkens,Ciobanu}. This approximation, valid if the numbers of bosons $N=\sum_{j=1} N_j$ and $M=\sum_{j=1} M_j$ are large enough, allows one to associate a certain Fock state $\ket{\vec{N},\vec{M}}=\ket{N_1,\dots,N_L,M_1,\dots,M_L}$ to state $\ket{\vec{x},\vec{y}}=\ket{x_1,\dots,x_L,y_1,\dots,y_L}$, i.e. to identify quantum numbers $N_j,\,M_j\in\mathbb{N}$ with labels $x_j,\,y_j\,\in[0,1]$. In this perspective, creation and annihilation processes $N_j\to N_j\pm 1$ ($M_j\to M_j\pm 1$) correspond to small variations $x_j\to x_j\pm\epsilon_a$ ($y_j\to y_j\pm\epsilon_b$), where $\epsilon_a=1/N\ll 1$ ($\epsilon_b=1/M\ll 1$). This scheme leads to a new effective Hamiltonian written in terms of coordinates $x_j$, $y_j$ and of their generalized conjugate momenta \cite{BuonsantePennaVezzani}. Accordingly, in the CVP framework, the relevant eigenvalue problem $\hat{H}\ket{E}=E\ket{E}$ (after expanding quanity $\hat{H}\ket{E}$ up to the second order) can be recast as
\begin{equation}
\label{eq:Eigenvalue_problem}
 -(D+V)\psi_E(\vec{x},\vec{y}) = E\,\psi_E(\vec{x},\vec{y}).
\end{equation}
The application of this rather versatile approximation scheme to Hamiltonian  (\ref{eq:Hamiltoniana}) (which describes a binary mixture confined in a three-well optical lattice with periodic boundary conditions) leads to an effective eigenvalue problem of the type (\ref{eq:Eigenvalue_problem}) where
$$
  D=  -\frac{T_a}{N} \sum_{j=1}^3 \left[\left(\frac{\partial}{\partial x_j}-\frac{\partial}{\partial x_{j+1}}\right) \sqrt{x_j x_{j+1}}\left(\frac{\partial}{\partial x_j}-\frac{\partial}{\partial x_{j+1}}\right)  \right] 
  -\frac{T_b}{M} \sum_{j=1}^3 \left[\left(\frac{\partial}{\partial y_j}-\frac{\partial}{\partial y_{j+1}}\right) \sqrt{y_j y_{j+1}}\left(\frac{\partial}{\partial y_j}-\frac{\partial}{\partial y_{j+1}}\right)  \right]
$$
is the generalized Laplacian and 
$$
   V= -2NT_a\left(\sqrt{x_1x_2}+\sqrt{x_2x_3}+\sqrt{x_3x_1}\right)-2MT_b\left(\sqrt{y_1y_2}+\sqrt{y_2y_3}+\sqrt{y_3y_1}\right)
$$
$$
  +\frac{U_a N^2}{2} \left[x_1(x_1-\epsilon_a)+x_2(x_2-\epsilon_a)+x_3(x_3-\epsilon_a) \right]
  +\frac{U_b M^2}{2} \left[y_1(y_1-\epsilon_b)+y_2(y_2-\epsilon_b)+y_3(y_3-\epsilon_b) \right]
$$
$$
  + W NM\left[x_1y_1 +x_2y_2 + x_3 y_3\right]
$$
is the generalized potential. The latter can be used to obtain meaningful information about the ground-state structure as a function of the model parameters. Notice that, in the CVP framework, expressions $N=\sum_{j=1}^3 N_j$ and $M=\sum_{j=1}^3 M_j$, giving the particle numbers conservation, read $1=\sum_{i=1}^3 x_i$ and $1=\sum_{j=1}^3 y_j$ respectively. This circumstance entails that the two terms proportional to $\epsilon_a$ and $\epsilon_b$ represent constant quantities. Effective potential (\ref{eq:V_*}) is obtained by choosing $T_a=T_b=:T$, $U_a=U_b=:U$ and $N=M$.

\begin{figure}[h!]
\includegraphics[width=1.0\columnwidth]{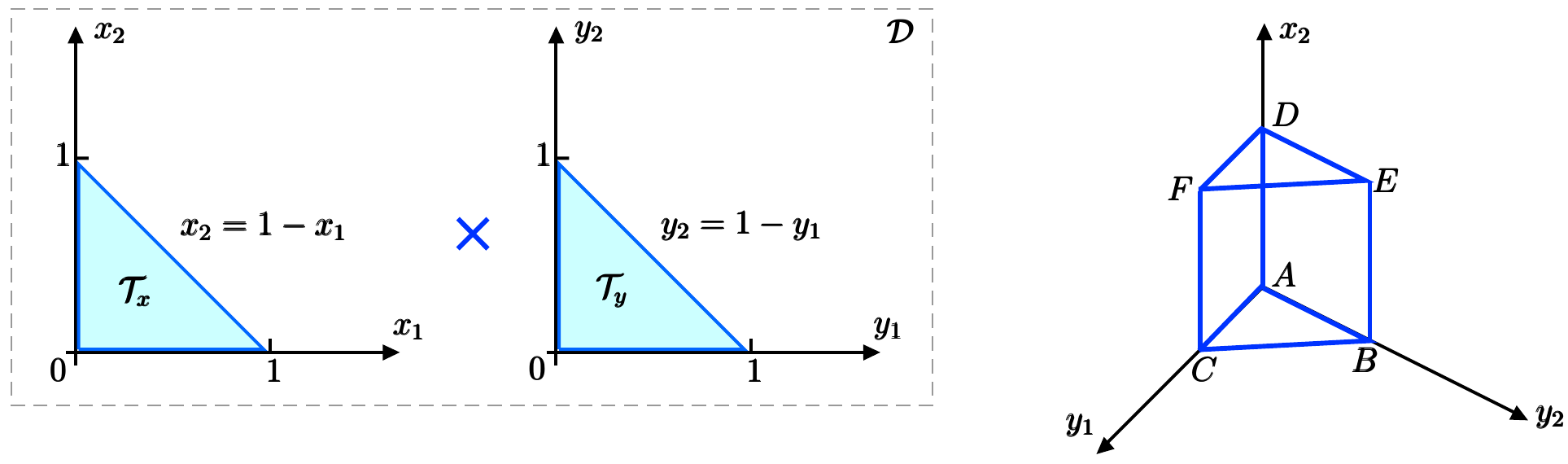}
\centering
\caption{Left panel: the domain of function $\mathcal{V}$ is the direct product of two triangular bi-dimensional regions. Right panel: one of the three-dimensional objects which constitutes the boundary of the 4-polytope $\mathcal{D}$. }
\label{fig:Dominio_4D}
\end{figure}

\paragraph{Minimum-energy configuration for $\pmb{T =0}$: geometric approach on a polytope-like domain.} If the tunnelling is suppressed, effective potential (\ref{eq:V_*}) simplifies as follows
\begin{equation}
    \label{eq:V_T_nullo}
    \mathcal{V}=  \frac{UN^2}{2}\left(x_1^2+x_2^2+x_3^2+ y_1^2+y_2^2+y_3^2 \right)+ WN^2 \left(x_1y_1+x_2y_2+x_3 y_3 \right),
\end{equation}
where $x_3=1-x_1-x_2$ and $y_3=1-y_1-y_2$ are, actually, dependent variables. The minimum-energy configuration corresponds to the global minimum of $\mathcal{V}(x_1,x_2,y_1,y_2)$, which is a function
$ \mathcal{V}:\mathcal{D} \subset \mathbb{R}^4  \to  \mathbb{R}$.
The domain $\mathcal{D}$ is a 4-polytope, consisting in the direct product of two triangular regions in $\mathbb{R}^2$, as shown in Fig. \ref{fig:Dominio_4D} (left panel). 
Computing the gradient with respect to the four independent variables, one finds that the only stationary point is the one associated to the uniform configuration, i.e $x_1=x_2=y_1=y_2=1/3$, corresponding to
\begin{equation}
\label{eq:V_uniform}
\mathcal{V}\left(\frac{1}{3},\frac{1}{3},\frac{1}{3},\frac{1}{3}\right)=\frac{N^2}{3}(U+W).
\end{equation}
The eigenvalues of the associated Hessian matrix show that this stationary point is a minimum provided that $W<U$. Notice that, in general, computing the stationary points does not necessarily give the \textit{global} minimum. The latter, in fact, can live on the boundary of the domain $\mathcal{D}$, in a point where the four-dimensional gradient is not well defined. So, one has to compare the value of $\mathcal{V}$ at the local minimum, i.e. $N^2(U+W)/3$, with the values of $\mathcal{V}$ at the boundary of $\mathcal{D}$. 

The exhaustive exploration of the domain boundary is a fortiori necessary when $W>U$ i.e. when the stationary point $x_1=x_2=y_1=y_2=1/3$ is no longer a minimum, a circumstance which entails that the global minimum lives on the domain boundary.
The complexity of our problem is due to the fact that the domain of $\mathcal{V}$ is four-dimensional and so its boundary is the union of six three-dimensional objects of the type sketched in Fig. \ref{fig:Dominio_4D} (right panel). Each of them corresponds to a Cartesian product $I_j^\prime \times \mathcal{T}_x$ and $I_j \times \mathcal{T}_y$ (where $I_j$  and $I_j^\prime$, with $j=1,\,2,\,3$ are the edges of triangles $\mathcal{T}_x$  and $\mathcal{T}_y$, respectively) and constitutes the domain of a 3-variable function obtained by introducing an additional constraint in formula (\ref{eq:V_T_nullo}). In the same spirit, the global minimum of these constrained functions must be searched not only in the interior of their domain (imposing the stationarity of a three-dimensional gradient), but also on their boundaries, by means of an exhaustive exploration thereof. One therefore iterates this process looking for stationary points, at first inside the \textit{volume}, then on the \textit{faces}, on the \textit{edges} (employing, at each step, a lower-dimensional gradient) and, eventually, evaluates the function at the \textit{vertices} of this 3D region.
The resulting set of candidates for the global minimum is such that each local minimum is linked to an existence condition (typically, an inequality in the one-dimensional space $W/U$) on the sub-domain where it was found. The global minimum, in each interval of the space $W/U$, is determined by comparing all the possible candidates. In the following we sketch the application of this rather general scheme to our problem.

We start fixing $x_1=0$. Notice that the other possible ways of fixing the first variable, i.e. $x_2=0$, $x_2=-x_1+1$, $y_1=0$, $y_2=0$ and $y_2=-y_1+1$ would lead to minimum-energy configurations which are equivalent up to cyclic permutations of the indexes and/or species labels swapping. The resulting constrained function
$$
 \mathcal{V}_3= \mathcal{V}(x_2,\,y_1,\,y_2\,;\, x_1=0) = UN^2(1-x_2 +x_2^2-y_1+y_1^2-y_2+y_1y_2+y_2^2) + WN^2(-x_2+y_1+x_2y_1+2x_2y_2)
$$
is defined on the \textit{prism} represented in the right panel of Fig. \ref{fig:Dominio_4D}. The stationary point of this 3-variables function, computed imposing $\nabla_{x_2,y_1,y_2} \mathcal{V}_{3} =(0,0,0)$ is a local minimum of $\mathcal{V}_3$ if $W<U$ but it must be discarded because the corresponding value of $\mathcal{V}$ is greater than (\ref{eq:V_uniform}). Let us focus on the $5$ \textit{faces} of this prism. Each face is fixed by introducing an additional constraint. As an example, face ABC is the domain of function
 $$
  \mathcal{V}_2 =    \mathcal{V}(y_1,\,y_2\,;\, x_1=0,\, x_2=0) = UN^2(1-y_1+y_1^2 -y_2 + y_1y_2 +y_2^2) +WN^2(1-y_1-y_2)
    $$
Computing the gradient, $\nabla_{y_1,y_2}$ one finds that there is one stationary point, which is $\left( y_1 = \frac{U+W}{3U},\, y_2 = \frac{U+W}{3U} \right)$. This is a minimum (both eigenvalues of the Hessian matrix bigger than 0) in the domain of interest iff $W<U/2$. Nevertheless, the corresponding value of $\mathcal{V}$ is smaller than the uniform-configuration value (\ref{eq:V_uniform}) just for $W>U$. But, in this range, this stationary point is no longer a minimum, so it must be discarded. The analysis of the $5$ faces, carried out according to this scheme, leads to the conclusion that, for $1<W/U<2$, there are $2$ minimum-energy configurations
$$
\left(x_1 =0,\quad x_2= \frac{U+W}{2U+W},\quad x_3 = \frac{U}{2U+W},\quad y_1 = \frac{U+W}{2U+W},\quad y_2=0,\quad y_3= \frac{U}{2U+W}  \right) \quad \leftarrow \quad \text{Face ACFD}
$$
and 
$$
\left(x_1 =0,\quad x_2= \frac{U}{2U+W},\quad x_3 = \frac{U+W}{2U+W},\quad y_1 = \frac{U+W}{2U+W},\quad y_2=\frac{U}{2U+W},\quad y_3= 0 \right) \quad \leftarrow \quad \text{Face CBEF}.
$$
Notice that these two configurations are equivalent up to a permutation of site indexes. The first one has been plotted in Fig. \ref{fig:Popolazioni_T_nullo}. 

The $9$ \textit{edges} of the solid of Fig. \ref{fig:Dominio_4D} (right panel) are identified by fixing an additional constraint. Their analysis shows that local iso-energetic points of minimum are present in segments CB DF and CF. Such minimum energy configurations are equal up to permutation of site indexes and/or species labels swapping. For example, segment CB is the domain of function 
$$
  \mathcal{V}_1 =   \mathcal{V}(y_1\,;\,x_1=0,\, x_2=0,\,y_2=1-y_1) = UN^2(1-y_1+y_1^2).
$$
Computing the derivative, $\frac{d}{d\,y_1}$, one finds that there is only one stationary point, which is $y_1= \frac{1}{2}$. Computing the second order derivative one can verify that this stationary point is always a local minimum in the physical interval $[0,1]$. The corresponding value $\frac{3}{4}UN^2$ of $\mathcal{V}$, is less than the one found in faces ACFD and CBEF provided that $W>2U$. Eventually, the analysis of the $6$ \textit{vertices} does not give any further minimum-energy configuration.

\paragraph{The search for the minimum-energy configuration for $T\neq0$.} The minimum-energy configuration is the configuration $(\vec{x},\, \vec{y})=(x_1,\,x_2,\,x_3,\,y_1,\,y_2,\,y_3)$ which minimizes effective potential (\ref{eq:V_*}). The presence of a non-vanishing $T$ constitutes a regularizing term which makes the exploration of the domain boundary no longer necessary. This fact, verified numerically, can be understood in physical terms because the presence of $T\neq 0$ makes configurations where one or more populations are exactly zero impossible. To compute the minimum-energy configuration, one therefore imposes stationary conditions for effective potential $V_*$ in which $x_3=1-x_1-x_2$ and $y_3=1-y_1-y_2$ take into account the boson number conservation. The resulting system of equations is 
\begin{equation}
\label{eq:Grad_nullo}
   \left(\frac{\partial V_*}{\partial x_1}, \, \frac{\partial V_*}{\partial x_2}, \,\frac{\partial V_*}{\partial y_1}, \,\frac{\partial V_*}{\partial y_2} \right) =(0,0,0,0)
\end{equation}
and the explicit form thereof is
\begin{equation}
\label{eq:Sistema_4}
\begin{cases}
 -2 N T \left(-\frac{x_2}{\sqrt{-x_2 (x_1+x_2-1)}}+\sqrt{\frac{x_2}{x_1}}-\frac{2 x_1+x_2-1}{\sqrt{-x_1 (x_1+x_2-1)}}\right)+2 UN^2 (2 x_1+x_2-1)+2 WN^2 (2 y_1+y_2-1) =0 
\\
 -2 N T \left(-\frac{x_1}{\sqrt{-x_1 (x_1+x_2-1)}}+\sqrt{\frac{x_1}{x_2}}-\frac{x_1+2 x_2-1}{\sqrt{-x_2 (x_1+x_2-1)}}\right)+2 UN^2 (x_1+2 x_2-1)+2 WN^2 (y_1+2 y_2-1) =0
\\
 -2 N T \left(-\frac{y_2}{\sqrt{-y_2 (y_1+y_2-1)}}+\sqrt{\frac{y_2}{y_1}}-\frac{2 y_1+y_2-1}{\sqrt{-y_1 (y_1+y_2-1)}}\right)+2UN^2 (2 y_1+y_2-1)+2 WN^2 (2 x_1+x_2-1) =0
\\
  -2 N T \left(-\frac{y_1}{\sqrt{-y_1 (y_1+y_2-1)}}+\sqrt{\frac{y_1}{y_2}}-\frac{y_1+2 y_2-1}{\sqrt{-y_2 (y_1+y_2-1)}}\right)+2 UN^2 (y_1+2 y_2-1)+2 WN^2 (x_1+2 x_2-1) =0.
  \end{cases}
\end{equation}
In general, it is not possible to find all the possible solutions of this system in a closed form and one needs to resort to numerical methods. Boson populations shown in Fig. \ref{fig:T_non_nulli_4} not only fulfill system (\ref{eq:Sistema_4}), but have been checked (by evaluating the eigenvalues of each associated Hessian matrix), to be minimum points of the effective potential (\ref{eq:V_*}). As already mentioned, Fig. \ref{fig:T_non_nulli_4} illustrates a particular \textit{family} of solutions, meaning that those configurations which are obtained by means of cyclic permutations of site indexes and/or by species-labels swapping, are still solutions of system (\ref{eq:Sistema_4}) and therefore points of minimum of (\ref{eq:V_*}). 

It is worth noticing that a particularly simple and significant solution of system (\ref{eq:Sistema_4}) is the one which corresponds to the uniform configuration $x_1=x_2=y_1=y_2=1/3$. This stationary point is a minimum of (\ref{eq:V_*}) provided that the associated Hessian matrix
\begin{equation}
\label{eq:Hess}
    \pmb{H}=N^2 \left(
\begin{array}{cccc}
 \frac{9 T}{N}+2 U & \frac{9 T}{2 N}+U & 2 W & W \\
 \\
 \frac{9 T}{2 N}+U & \frac{9 T}{N}+2 U & W & 2 W \\
 \\
 2 W & W & \frac{9 T}{N}+2 U & \frac{9 T}{2 N}+U \\
 \\
 W & 2 W & \frac{9 T}{2 N}+U & \frac{9 T}{N}+2 U \\
\end{array}
\right)
\end{equation}
is definite positive, a condition which is verified iff $W/U<1+(9T)/(2UN)$. This argument proves the correctness of critical value (\ref{eq:critical_ration}), presented in the main text and corresponding to purple solid lines in Fig. \ref{fig:Diagramma_di_Fase}.

An effective way to reduce the complexity of system (\ref{eq:Grad_nullo}) (and possibly simplify its numerical solution) comes from the symmetry which marks the boson populations when the system is in phase (ii) (see central regions of Fig. \ref{fig:T_non_nulli_4} panels). In such phase, i.e. for moderate $W/U$ values, one can notice, in fact, that $y_2=x_1$ and $y_1=x_2$, two constraints that allow to rewrite system (\ref{eq:Grad_nullo}) as
\begin{equation}
    \label{eq:System_2_equations}
\begin{cases}
  -2 N T \left(-\frac{x_2}{\sqrt{-x_2 (x_1+x_2-1)}}+\sqrt{\frac{x_2}{x_1}}-\frac{2 x_1+x_2-1}{\sqrt{-x_1 (x_1+x_2-1)}}\right)+2 UN^2 (2 x_1+x_2-1)+2 WN^2 (x_1+2 x_2-1) = 0
\\
  -2 N T \left(-\frac{x_1}{\sqrt{-x_1 (x_1+x_2-1)}}+\sqrt{\frac{x_1}{x_2}}-\frac{x_1+2 x_2-1}{\sqrt{-x_2 (x_1+x_2-1)}}\right)+2 UN^2 (x_1+2 x_2-1)+2 WN^2 (2 x_1+x_2-1) = 0.
\end{cases}
\end{equation}
The orange dashed lines in Fig. \ref{fig:Diagramma_di_Fase}, which constitute the border between the intermediate and the fully demixed phase, have no analytical expression and correspond to those solutions of system (\ref{eq:System_2_equations}) which are no longer minimum points for effective potential (\ref{eq:V_*}), i.e. to stationary points for which at least one Hessian-matrix eigenvalue vanishes.  
Incidentally, critical condition (\ref{eq:critical_ration}), obtained ``from the left" by means of Hessian matrix (\ref{eq:Hess}), can be obtained ``from the right" also. This can be understood by linearizing equations (\ref{eq:System_2_equations}) around the known solution $\left(x_1=\frac{1}{3}, x_2=\frac{1}{3}\right)$. Substituting $x_1=\frac{1}{3}-\epsilon_1$ and $x_2=\frac{1}{3}+\epsilon_2$ and expanding up to the first order for $\epsilon_1 \to 0$  and $\epsilon_2 \to 0 $, one gets
$$
\begin{cases}
 \epsilon_1 (-9 N T-2 UN^2-WN^2)+\epsilon_2 \left(\frac{9 N T}{2}+UN^2+2 WN^2\right)=0
\\
  \epsilon_1 \left(-\frac{1}{2} 9 N T-UN^2-2 WN^2\right)+\epsilon_2 (9 N T+2 UN^2+WN^2)=0.
\end{cases}
$$
This linear system admits non-trivial solutions provided that the determinant is zero, a condition which leads, again, to critical value (\ref{eq:critical_ration}).

\paragraph{Entanglement Entropy.}
Let us consider the system ground state, $\ket{\psi_0}$ and let us choose, as a basis, the set $\ket{n_1,n_2,N-n_1-n_2,m_1,m_2,M-m_1-m_2}$ (notice that, for the sake of clarity, $N$ ($M$) indicates the total number of bosons of species a (b)). It is therefore possible to expand $\ket{\psi_0}$ with respect to this basis:
$$
  \ket{\psi_0}=\sum_{i=0}^N \sum_{j=0}^{N-i}\sum_{k=0}^M\sum_{l=0}^{M-k}\,\, C_{i,j,k,l}\, \ket{i,j,N-i-j,k,l,M-k-l}
$$
where
$$
   C_{i,j,k,l} = \braket{i,j,N-i-j,k,l,M-k-l}{\psi_0}.
$$
As a consequence, the density matrix associated to the system ground state reads:
$$
   \hat{\rho}_0=  \ket{\psi_0}\bra{\psi_0} = 
$$
$$
   =\sum_{i=0}^N \sum_{j=0}^{N-i}\sum_{k=0}^M\sum_{l=0}^{M-k}
   \sum_{i^\prime=0}^N \sum_{j^\prime=0}^{N-i^\prime}\sum_{k^\prime=0}^M\sum_{l^\prime=0}^{M-k^\prime}
  C_{i,j,k,l} \, C_{i^\prime, j^\prime, k^\prime, l^\prime}^* \,\, \ket{i,j,N-i-j,k,l,M-k-l}  \bra{i^\prime,j^\prime ,N-i^\prime -j^\prime,k^\prime,l^\prime,M-k^\prime-l^\prime}.
$$ 
This is the density matrix of a \textit{pure state} and so one can check that it is hermitian, it has unitary trace, it is equal to its square, and its eigenvalues are all zero except for one which is $1$.

A possible and quite natural way to sub-divide the system into two partitions leads to consider the entanglement between the two condensed species. The basis relevant to the sub-system ``species A" is $\{\ket{n_1,n_2,N-n_1-n_2}\}$ while the basis relevant to the sub-system ``species B" is $\{\ket{m_1,m_2,M-m_1-m_2}\}$. It is possible to compute the reduced density matrix relevant to the ``species A" by tracing out the degrees of freedom relevant to ``species B"
\begin{equation}
\label{eq:Reduced_density_matrix}
 \hat{\rho}_{0,a} = \mathrm{Tr}_b\left(\hat{\rho}_{0}\right)= \sum_{p=0}^M\sum_{q=0}^{M-p} \bra{p,q,M-p-q} \,\, \hat{\rho}_0 \,\, \ket{p,q,M-p-q}=
\end{equation}
$$
 =\sum_{i=0}^N \sum_{j=0}^{N-i}
   \sum_{i^\prime=0}^N \sum_{j^\prime=0}^{N-i^\prime}
  M_{i,j,i^\prime,j^\prime} 
  \ket{i,j,N-i-j}  \,\,\,  \bra{i^\prime,j^\prime ,N-i^\prime -j^\prime}
$$
where 
$$
   M_{i,j,i^\prime,j^\prime}  = \sum_{k=0}^M \sum_{l=0}^{M-k}  C_{i,j,k,l} \, C_{i^\prime, j^\prime, k, l}^*.
$$
This is the reduced density matrix of a  \textit{mixed state}, whose Von Neumann entropy
\begin{equation}
\label{eq:Von_Neumann_Entropy}
  EE=-\mathrm{Tr}(\hat{\rho}_{0,a} \, \log_2 \hat{\rho}_{0,a} )=-\sum_{j=0} \lambda_j \,\log_2 \lambda_j
\end{equation}
(where $\lambda_j's$ are the eigenvalues of $\hat{\rho}_{0,a}$) corresponds to the EE between the two parts of the global system.

\section*{Author contributions statement}
A.R. performed analytic and numerical calculations.  A.R. and V.P. have analyzed the results and equally contributed in writing and reviewing the manuscript. V.P. supervised the work. 

\section*{Additional information}
\textbf{Competing interests:} The authors declare no competing interests.  



\begin{thebibliography}{1}
\expandafter\ifx\csname url\endcsname\relax
  \def\url#1{\texttt{#1}}\fi
\expandafter\ifx\csname urlprefix\endcsname\relax\def\urlprefix{URL }\fi
\expandafter\ifx\csname doiprefix\endcsname\relax\def\doiprefix{DOI }\fi
\providecommand{\bibinfo}[2]{#2}
\providecommand{\eprint}[2][]{\url{#2}}

\bibitem{Figueredo:2009dg}
\bibinfo{author}{Figueredo, A.~J.} \& \bibinfo{author}{Wolf, P. S.~A.}
\newblock \bibinfo{journal}{\bibinfo{title}{Assortative pairing and life
  history strategy - a cross-cultural study.}}
\newblock {\emph{\JournalTitle{Human Nature}}} \textbf{\bibinfo{volume}{20}},
  \bibinfo{pages}{317--330} (\bibinfo{year}{2009}).

\end{thebibliography}


\begin{thebibliography}{10}
\expandafter\ifx\csname url\endcsname\relax
  \def\url#1{\texttt{#1}}\fi
\expandafter\ifx\csname urlprefix\endcsname\relax\def\urlprefix{URL }\fi
\expandafter\ifx\csname doiprefix\endcsname\relax\def\doiprefix{DOI }\fi
\providecommand{\bibinfo}[2]{#2}
\providecommand{\eprint}[2][]{\url{#2}}

\bibitem{cmixt1}
\bibinfo{author}{Ho, T.-L.} \& \bibinfo{author}{Shenoy, V.~B.}
\newblock \bibinfo{journal}{\bibinfo{title}{Binary mixtures of {B}ose
  condensates of alkali atoms}}.
\newblock {\emph{\JournalTitle{Phys. Rev. Lett.}}}
  \textbf{\bibinfo{volume}{77}}, \bibinfo{pages}{3276--3279}
  (\bibinfo{year}{1996}).

\bibitem{cmixt2}
\bibinfo{author}{Pu, H.} \& \bibinfo{author}{Bigelow, N.~P.}
\newblock \bibinfo{journal}{\bibinfo{title}{Properties of two-species {B}ose
  condensates}}.
\newblock {\emph{\JournalTitle{Phys. Rev. Lett.}}}
  \textbf{\bibinfo{volume}{80}}, \bibinfo{pages}{1130--1133}
  (\bibinfo{year}{1998}).

\bibitem{cmixt3}
\bibinfo{author}{\"Ohberg, P.} \& \bibinfo{author}{Stenholm, S.}
\newblock \bibinfo{journal}{\bibinfo{title}{Hartree-fock treatment of the
  two-component {B}ose-{E}instein condensate}}.
\newblock {\emph{\JournalTitle{Phys. Rev. A}}} \textbf{\bibinfo{volume}{57}},
  \bibinfo{pages}{1272--1279} (\bibinfo{year}{1998}).

\bibitem{cmixt4}
\bibinfo{author}{Esry, B.~D.} \& \bibinfo{author}{Greene, C.~H.}
\newblock \bibinfo{journal}{\bibinfo{title}{Spontaneous spatial symmetry
  breaking in two-component {B}ose-{E}instein condensates}}.
\newblock {\emph{\JournalTitle{Phys. Rev. A}}} \textbf{\bibinfo{volume}{59}},
  \bibinfo{pages}{1457--1460} (\bibinfo{year}{1999}).

\bibitem{cmixt5}
\bibinfo{author}{Svidzinsky, A.~A.} \& \bibinfo{author}{Chui, S.~T.}
\newblock \bibinfo{journal}{\bibinfo{title}{Symmetric-asymmetric transition in
  mixtures of {B}ose-{E}instein condensates}}.
\newblock {\emph{\JournalTitle{Phys. Rev. A}}} \textbf{\bibinfo{volume}{67}},
  \bibinfo{pages}{053608} (\bibinfo{year}{2003}).

\bibitem{kasa}
\bibinfo{author}{Kasamatsu, K.} \& \bibinfo{author}{Tsubota, M.}
\newblock \bibinfo{journal}{\bibinfo{title}{Modulation instability and
  solitary-wave formation in two-component {B}ose-{E}instein condensates}}.
\newblock {\emph{\JournalTitle{Phys. Rev. A}}} \textbf{\bibinfo{volume}{74}},
  \bibinfo{pages}{013617} (\bibinfo{year}{2006}).

\bibitem{Gallemi_1}
\bibinfo{author}{Melé-Messeguer, M.}, \bibinfo{author}{Juliá-Díaz, B.},
  \bibinfo{author}{Guilleumas, M.}, \bibinfo{author}{Polls, A.} \&
  \bibinfo{author}{Sanpera, A.}
\newblock \bibinfo{journal}{\bibinfo{title}{Weakly linked binary mixtures of
  ${F}=1$ $^{87}${R}b {B}ose–{E}instein condensates}}.
\newblock {\emph{\JournalTitle{New Journal of Physics}}}
  \textbf{\bibinfo{volume}{13}}, \bibinfo{pages}{033012}
  (\bibinfo{year}{2011}).

\bibitem{tic}
\bibinfo{author}{Ticknor, C.}
\newblock \bibinfo{journal}{\bibinfo{title}{Excitations of a trapped
  two-component {B}ose-{E}instein condensate}}.
\newblock {\emph{\JournalTitle{Phys. Rev. A}}} \textbf{\bibinfo{volume}{88}},
  \bibinfo{pages}{013623} (\bibinfo{year}{2013}).

\bibitem{lee}
\bibinfo{author}{Lee, K.~L.} \emph{et~al.}
\newblock \bibinfo{journal}{\bibinfo{title}{Phase separation and dynamics of
  two-component {B}ose-{E}instein condensates}}.
\newblock {\emph{\JournalTitle{Phys. Rev. A}}} \textbf{\bibinfo{volume}{94}},
  \bibinfo{pages}{013602} (\bibinfo{year}{2016}).

\bibitem{jz}
\bibinfo{author}{Jaksch, D.} \& \bibinfo{author}{Zoller, P.}
\newblock \bibinfo{journal}{\bibinfo{title}{The cold atom {H}ubbard toolbox}}.
\newblock {\emph{\JournalTitle{Annals of Physics}}}
  \textbf{\bibinfo{volume}{315}}, \bibinfo{pages}{52 -- 79}
  (\bibinfo{year}{2005}).
\newblock \bibinfo{note}{Special Issue}.

\bibitem{bdz}
\bibinfo{author}{Bloch, I.}, \bibinfo{author}{Dalibard, J.} \&
  \bibinfo{author}{Zwerger, W.}
\newblock \bibinfo{journal}{\bibinfo{title}{Many-body physics with ultracold
  gases}}.
\newblock {\emph{\JournalTitle{Rev. Mod. Phys.}}}
  \textbf{\bibinfo{volume}{80}}, \bibinfo{pages}{885--964}
  (\bibinfo{year}{2008}).

\bibitem{yuk}
\bibinfo{author}{Yukalov, V.~I.}
\newblock \bibinfo{journal}{\bibinfo{title}{Cold bosons in optical lattices}}.
\newblock {\emph{\JournalTitle{Laser Physics}}} \textbf{\bibinfo{volume}{19}},
  \bibinfo{pages}{1--110} (\bibinfo{year}{2009}).

\bibitem{Inguscio}
\bibinfo{author}{Thalhammer, G.} \emph{et~al.}
\newblock \bibinfo{journal}{\bibinfo{title}{Double species {B}ose-{E}instein
  condensate with tunable interspecies interactions}}.
\newblock {\emph{\JournalTitle{Phys. Rev. Lett.}}}
  \textbf{\bibinfo{volume}{100}}, \bibinfo{pages}{210402}
  (\bibinfo{year}{2008}).

\bibitem{Gadway}
\bibinfo{author}{Gadway, B.}, \bibinfo{author}{Pertot, D.},
  \bibinfo{author}{Reimann, R.} \& \bibinfo{author}{Schneble, D.}
\newblock \bibinfo{journal}{\bibinfo{title}{Superfluidity of interacting
  bosonic mixtures in optical lattices}}.
\newblock {\emph{\JournalTitle{Phys. Rev. Lett.}}}
  \textbf{\bibinfo{volume}{105}}, \bibinfo{pages}{045303}
  (\bibinfo{year}{2010}).

\bibitem{Soltan}
\bibinfo{author}{Soltan-Panahi, P.}, \bibinfo{author}{L{\"u}hmann, D.-S.},
  \bibinfo{author}{Struck, J.}, \bibinfo{author}{Windpassinger, P.} \&
  \bibinfo{author}{Sengstock, K.}
\newblock \bibinfo{journal}{\bibinfo{title}{Quantum phase transition to
  unconventional multi-orbital superfluidity in optical lattices}}.
\newblock {\emph{\JournalTitle{Nature Physics}}} \textbf{\bibinfo{volume}{8}},
  \bibinfo{pages}{71} (\bibinfo{year}{2012}).

\bibitem{sep1}
\bibinfo{author}{Mishra, T.}, \bibinfo{author}{Pai, R.~V.} \&
  \bibinfo{author}{Das, B.~P.}
\newblock \bibinfo{journal}{\bibinfo{title}{Phase separation in a two-species
  {B}ose mixture}}.
\newblock {\emph{\JournalTitle{Phys. Rev. A}}} \textbf{\bibinfo{volume}{76}},
  \bibinfo{pages}{013604} (\bibinfo{year}{2007}).

\bibitem{sep2}
\bibinfo{author}{Jain, P.} \& \bibinfo{author}{Boninsegni, M.}
\newblock \bibinfo{journal}{\bibinfo{title}{Quantum demixing in binary mixtures
  of dipolar bosons}}.
\newblock {\emph{\JournalTitle{Phys. Rev. A}}} \textbf{\bibinfo{volume}{83}},
  \bibinfo{pages}{023602} (\bibinfo{year}{2011}).

\bibitem{sep3}
\bibinfo{author}{Lingua, F.}, \bibinfo{author}{Guglielmino, M.},
  \bibinfo{author}{Penna, V.} \& \bibinfo{author}{Capogrosso~Sansone, B.}
\newblock \bibinfo{journal}{\bibinfo{title}{Demixing effects in mixtures of two
  bosonic species}}.
\newblock {\emph{\JournalTitle{Phys. Rev. A}}} \textbf{\bibinfo{volume}{92}},
  \bibinfo{pages}{053610} (\bibinfo{year}{2015}).

\bibitem{Angom}
\bibinfo{author}{Suthar, K.} \& \bibinfo{author}{Angom, D.}
\newblock \bibinfo{journal}{\bibinfo{title}{Optical-lattice-influenced geometry
  of quasi-two-dimensional binary condensates and quasiparticle spectra}}.
\newblock {\emph{\JournalTitle{Phys. Rev. A}}} \textbf{\bibinfo{volume}{93}},
  \bibinfo{pages}{063608} (\bibinfo{year}{2016}).

\bibitem{ks}
\bibinfo{author}{Kuklov, A.~B.} \& \bibinfo{author}{Svistunov, B.~V.}
\newblock \bibinfo{journal}{\bibinfo{title}{Counterflow superfluidity of
  two-species ultracold atoms in a commensurate optical lattice}}.
\newblock {\emph{\JournalTitle{Phys. Rev. Lett.}}}
  \textbf{\bibinfo{volume}{90}}, \bibinfo{pages}{100401}
  (\bibinfo{year}{2003}).

\bibitem{ddl}
\bibinfo{author}{Duan, L.-M.}, \bibinfo{author}{Demler, E.} \&
  \bibinfo{author}{Lukin, M.~D.}
\newblock \bibinfo{journal}{\bibinfo{title}{Controlling spin exchange
  interactions of ultracold atoms in optical lattices}}.
\newblock {\emph{\JournalTitle{Phys. Rev. Lett.}}}
  \textbf{\bibinfo{volume}{91}}, \bibinfo{pages}{090402}
  (\bibinfo{year}{2003}).

\bibitem{qe1}
\bibinfo{author}{Roscilde, T.} \& \bibinfo{author}{Cirac, J.~I.}
\newblock \bibinfo{journal}{\bibinfo{title}{Quantum emulsion: A glassy phase of
  bosonic mixtures in optical lattices}}.
\newblock {\emph{\JournalTitle{Phys. Rev. Lett.}}}
  \textbf{\bibinfo{volume}{98}}, \bibinfo{pages}{190402}
  (\bibinfo{year}{2007}).

\bibitem{mott}
\bibinfo{author}{Guglielmino, M.}, \bibinfo{author}{Penna, V.} \&
  \bibinfo{author}{Capogrosso-Sansone, B.}
\newblock \bibinfo{journal}{\bibinfo{title}{Mott-insulator to superfluid
  transition in {B}ose-{B}ose mixtures in a two-dimensional lattice}}.
\newblock {\emph{\JournalTitle{Phys. Rev. A}}} \textbf{\bibinfo{volume}{82}},
  \bibinfo{pages}{021601} (\bibinfo{year}{2010}).

\bibitem{pol}
\bibinfo{author}{Benjamin, D.} \& \bibinfo{author}{Demler, E.}
\newblock \bibinfo{journal}{\bibinfo{title}{Variational polaron method for
  {B}ose-{B}ose mixtures}}.
\newblock {\emph{\JournalTitle{Phys. Rev. A}}} \textbf{\bibinfo{volume}{89}},
  \bibinfo{pages}{033615} (\bibinfo{year}{2014}).

\bibitem{ent}
\bibinfo{author}{Wang, W.}, \bibinfo{author}{Penna, V.} \&
  \bibinfo{author}{Capogrosso-Sansone, B.}
\newblock \bibinfo{journal}{\bibinfo{title}{Inter-species entanglement of
  {B}ose–{B}ose mixtures trapped in optical lattices}}.
\newblock {\emph{\JournalTitle{New Journal of Physics}}}
  \textbf{\bibinfo{volume}{18}}, \bibinfo{pages}{063002}
  (\bibinfo{year}{2016}).

\bibitem{modul}
\bibinfo{author}{Jin, G.-R.}, \bibinfo{author}{Kim, C.~K.} \&
  \bibinfo{author}{Nahm, K.}
\newblock \bibinfo{journal}{\bibinfo{title}{Modulational instability of
  two-component {B}ose-{E}instein condensates in an optical lattice}}.
\newblock {\emph{\JournalTitle{Phys. Rev. A}}} \textbf{\bibinfo{volume}{72}},
  \bibinfo{pages}{045601} (\bibinfo{year}{2005}).

\bibitem{NoiPRA2}
\bibinfo{author}{Penna, V.} \& \bibinfo{author}{Richaud, A.}
\newblock \bibinfo{journal}{\bibinfo{title}{Two-species boson mixture on a
  ring: A group-theoretic approach to the quantum dynamics of low-energy
  excitations}}.
\newblock {\emph{\JournalTitle{Phys. Rev. A}}} \textbf{\bibinfo{volume}{96}},
  \bibinfo{pages}{053631} (\bibinfo{year}{2017}).

\bibitem{PennaLinguaJPB}
\bibinfo{author}{Lingua, F.}, \bibinfo{author}{Mazzarella, G.} \&
  \bibinfo{author}{Penna, V.}
\newblock \bibinfo{journal}{\bibinfo{title}{Delocalization effects,
  entanglement entropy and spectral collapse of boson mixtures in a double
  well}}.
\newblock {\emph{\JournalTitle{Journal of Physics B: Atomic, Molecular and
  Optical Physics}}} \textbf{\bibinfo{volume}{49}}, \bibinfo{pages}{205005}
  (\bibinfo{year}{2016}).

\bibitem{PennaLinguaPRE}
\bibinfo{author}{Lingua, F.} \& \bibinfo{author}{Penna, V.}
\newblock \bibinfo{journal}{\bibinfo{title}{Continuous-variable approach to the
  spectral properties and quantum states of the two-component {B}ose-{H}ubbard
  dimer}}.
\newblock {\emph{\JournalTitle{Phys. Rev. E}}} \textbf{\bibinfo{volume}{95}},
  \bibinfo{pages}{062142} (\bibinfo{year}{2017}).

\bibitem{Gallemi_3}
\bibinfo{author}{De~Chiara, G.}, \bibinfo{author}{Lepori, L.},
  \bibinfo{author}{Lewenstein, M.} \& \bibinfo{author}{Sanpera, A.}
\newblock \bibinfo{journal}{\bibinfo{title}{Entanglement spectrum, critical
  exponents, and order parameters in quantum spin chains}}.
\newblock {\emph{\JournalTitle{Phys. Rev. Lett.}}}
  \textbf{\bibinfo{volume}{109}}, \bibinfo{pages}{237208}
  (\bibinfo{year}{2012}).

\bibitem{Gallemi_4}
\bibinfo{author}{Gallem\'{\i}, A.}, \bibinfo{author}{Guilleumas, M.},
  \bibinfo{author}{Mayol, R.} \& \bibinfo{author}{Sanpera, A.}
\newblock \bibinfo{journal}{\bibinfo{title}{Role of anisotropy in dipolar
  bosons in triple-well potentials}}.
\newblock {\emph{\JournalTitle{Phys. Rev. A}}} \textbf{\bibinfo{volume}{88}},
  \bibinfo{pages}{063645} (\bibinfo{year}{2013}).

\bibitem{PBF}
\bibinfo{author}{Buonsante, P.}, \bibinfo{author}{Franzosi, R.} \&
  \bibinfo{author}{Penna, V.}
\newblock \bibinfo{journal}{\bibinfo{title}{Control of unstable macroscopic
  oscillations in the dynamics of three coupled {B}ose condensates}}.
\newblock {\emph{\JournalTitle{Journal of Physics A: Mathematical and
  Theoretical}}} \textbf{\bibinfo{volume}{42}}, \bibinfo{pages}{285307}
  (\bibinfo{year}{2009}).

\bibitem{Jason}
\bibinfo{author}{Jason, P.} \& \bibinfo{author}{Johansson, M.}
\newblock \bibinfo{journal}{\bibinfo{title}{Quantum signatures of charge
  flipping vortices in the {B}ose-{H}ubbard trimer}}.
\newblock {\emph{\JournalTitle{Phys. Rev. E}}} \textbf{\bibinfo{volume}{94}},
  \bibinfo{pages}{052215} (\bibinfo{year}{2016}).

\bibitem{Bradly}
\bibinfo{author}{Bradly, C.~J.}, \bibinfo{author}{Rab, M.},
  \bibinfo{author}{Greentree, A.~D.} \& \bibinfo{author}{Martin, A.~M.}
\newblock \bibinfo{journal}{\bibinfo{title}{Coherent tunneling via adiabatic
  passage in a three-well {B}ose-{H}ubbard system}}.
\newblock {\emph{\JournalTitle{Phys. Rev. A}}} \textbf{\bibinfo{volume}{85}},
  \bibinfo{pages}{053609} (\bibinfo{year}{2012}).

\bibitem{Gallemi_2}
\bibinfo{author}{Gallemí, A.} \emph{et~al.}
\newblock \bibinfo{journal}{\bibinfo{title}{Fragmented condensation in
  {B}ose–{H}ubbard trimers with tunable tunnelling}}.
\newblock {\emph{\JournalTitle{New Journal of Physics}}}
  \textbf{\bibinfo{volume}{17}}, \bibinfo{pages}{073014}
  (\bibinfo{year}{2015}).

\bibitem{NoiEntropy}
\bibinfo{author}{Lingua, F.}, \bibinfo{author}{Richaud, A.} \&
  \bibinfo{author}{Penna, V.}
\newblock \bibinfo{journal}{\bibinfo{title}{Residual entropy and critical
  behavior of two interacting boson species in a double well}}.
\newblock {\emph{\JournalTitle{Entropy}}} \textbf{\bibinfo{volume}{20}},
  \bibinfo{pages}{84} (\bibinfo{year}{2018}).

\bibitem{RevModPhysAmico}
\bibinfo{author}{Amico, L.}, \bibinfo{author}{Fazio, R.},
  \bibinfo{author}{Osterloh, A.} \& \bibinfo{author}{Vedral, V.}
\newblock \bibinfo{journal}{\bibinfo{title}{Entanglement in many-body
  systems}}.
\newblock {\emph{\JournalTitle{Rev. Mod. Phys.}}}
  \textbf{\bibinfo{volume}{80}}, \bibinfo{pages}{517--576}
  (\bibinfo{year}{2008}).

\bibitem{BuonsantePennaVezzani}
\bibinfo{author}{Buonsante, P.}, \bibinfo{author}{Penna, V.} \&
  \bibinfo{author}{Vezzani, A.}
\newblock \bibinfo{journal}{\bibinfo{title}{Dynamical bifurcation as a
  semiclassical counterpart of a quantum phase transition}}.
\newblock {\emph{\JournalTitle{Phys. Rev. A}}} \textbf{\bibinfo{volume}{84}},
  \bibinfo{pages}{061601} (\bibinfo{year}{2011}).

\bibitem{Mazz}
\bibinfo{author}{Mazzarella, G.}, \bibinfo{author}{Salasnich, L.},
  \bibinfo{author}{Parola, A.} \& \bibinfo{author}{Toigo, F.}
\newblock \bibinfo{journal}{\bibinfo{title}{Coherence and entanglement in the
  ground state of a bosonic {J}osephson junction: From macroscopic
  {S}chr\"odinger cat states to separable {F}ock states}}.
\newblock {\emph{\JournalTitle{Phys. Rev. A}}} \textbf{\bibinfo{volume}{83}},
  \bibinfo{pages}{053607} (\bibinfo{year}{2011}).

\bibitem{DellAnnaMazzarella}
\bibinfo{author}{Dell'Anna, L.}, \bibinfo{author}{Mazzarella, G.},
  \bibinfo{author}{Penna, V.} \& \bibinfo{author}{Salasnich, L.}
\newblock \bibinfo{journal}{\bibinfo{title}{Entanglement entropy and
  macroscopic quantum states with dipolar bosons in a triple-well potential}}.
\newblock {\emph{\JournalTitle{Phys. Rev. A}}} \textbf{\bibinfo{volume}{87}},
  \bibinfo{pages}{053620} (\bibinfo{year}{2013}).

\bibitem{Hines}
\bibinfo{author}{Hines, A.~P.}, \bibinfo{author}{McKenzie, R.~H.} \&
  \bibinfo{author}{Milburn, G.~J.}
\newblock \bibinfo{journal}{\bibinfo{title}{Quantum entanglement and
  fixed-point bifurcations}}.
\newblock {\emph{\JournalTitle{Phys. Rev. A}}} \textbf{\bibinfo{volume}{71}},
  \bibinfo{pages}{042303} (\bibinfo{year}{2005}).

\bibitem{FuLiu}
\bibinfo{author}{Fu, L.} \& \bibinfo{author}{Liu, J.}
\newblock \bibinfo{journal}{\bibinfo{title}{Quantum entanglement manifestation
  of transition to nonlinear self-trapping for {B}ose-{E}instein condensates in
  a symmetric double well}}.
\newblock {\emph{\JournalTitle{Phys. Rev. A}}} \textbf{\bibinfo{volume}{74}},
  \bibinfo{pages}{063614} (\bibinfo{year}{2006}).

\bibitem{JuliaDiaz}
\bibinfo{author}{Juli\'a-D\'{\i}az, B.}, \bibinfo{author}{Dagnino, D.},
  \bibinfo{author}{Lewenstein, M.}, \bibinfo{author}{Martorell, J.} \&
  \bibinfo{author}{Polls, A.}
\newblock \bibinfo{journal}{\bibinfo{title}{Macroscopic self-trapping in
  {B}ose-{E}instein condensates: Analysis of a dynamical quantum phase
  transition}}.
\newblock {\emph{\JournalTitle{Phys. Rev. A}}} \textbf{\bibinfo{volume}{81}},
  \bibinfo{pages}{023615} (\bibinfo{year}{2010}).

\bibitem{Viscondi}
\bibinfo{author}{Viscondi, T.~F.}, \bibinfo{author}{Furuya, K.} \&
  \bibinfo{author}{de~Oliveira, M.~C.}
\newblock \bibinfo{journal}{\bibinfo{title}{Generalized purity and quantum
  phase transition for {B}ose-{E}instein condensates in a symmetric double
  well}}.
\newblock {\emph{\JournalTitle{Phys. Rev. A}}} \textbf{\bibinfo{volume}{80}},
  \bibinfo{pages}{013610} (\bibinfo{year}{2009}).

\bibitem{BuonsanteBurioni}
\bibinfo{author}{Buonsante, P.}, \bibinfo{author}{Burioni, R.},
  \bibinfo{author}{Vescovi, E.} \& \bibinfo{author}{Vezzani, A.}
\newblock \bibinfo{journal}{\bibinfo{title}{Quantum criticality in a bosonic
  {J}osephson junction}}.
\newblock {\emph{\JournalTitle{Phys. Rev. A}}} \textbf{\bibinfo{volume}{85}},
  \bibinfo{pages}{043625} (\bibinfo{year}{2012}).

\bibitem{PennaFranzosi2001}
\bibinfo{author}{Franzosi, R.} \& \bibinfo{author}{Penna, V.}
\newblock \bibinfo{journal}{\bibinfo{title}{Spectral properties of coupled
  {B}ose-{E}instein condensates}}.
\newblock {\emph{\JournalTitle{Phys. Rev. A}}} \textbf{\bibinfo{volume}{63}},
  \bibinfo{pages}{043609} (\bibinfo{year}{2001}).

\bibitem{Spekkens}
\bibinfo{author}{Spekkens, R.~W.} \& \bibinfo{author}{Sipe, J.~E.}
\newblock \bibinfo{journal}{\bibinfo{title}{Spatial fragmentation of a
  {B}ose-{E}instein condensate in a double-well potential}}.
\newblock {\emph{\JournalTitle{Phys. Rev. A}}} \textbf{\bibinfo{volume}{59}},
  \bibinfo{pages}{3868--3877} (\bibinfo{year}{1999}).

\bibitem{Ciobanu}
\bibinfo{author}{Ho, T.-L.} \& \bibinfo{author}{Ciobanu, C.~V.}
\newblock \bibinfo{journal}{\bibinfo{title}{The {S}chr{\"o}dinger cat family in
  attractive {B}ose gases}}.
\newblock {\emph{\JournalTitle{Journal of Low Temperature Physics}}}
  \textbf{\bibinfo{volume}{135}}, \bibinfo{pages}{257--266}
  (\bibinfo{year}{2004}).

\end{thebibliography}
\end{document}